\documentclass[aps,prl,reprint,twocolumn,superscriptaddress,showpacs]{revtex4-1}
\usepackage{amsmath,amssymb,graphicx,pdfpages}
\usepackage[colorlinks,hyperindex,linkcolor=blue,citecolor=blue,urlcolor=blue]{hyperref}
\newcommand{\figdraft}{false} 
\newcommand{\LBE}{\text{\tiny LBE}}
\newcommand{\EOS}{\text{\tiny EOS}}
\newcommand{\INT}{\text{\tiny INT}}
\newcommand{\Ma}{M\mspace{-3.0mu}a}
\newcommand{\We}{W\mspace{-3.0mu}e}
\newcommand{\Rey}{R\mspace{-1.0mu}e}

\begin{document}
\title{Lattice Boltzmann model with self-tuning equation of state for
multiphase flows}
\author{Rongzong Huang}
\email{rongzong.huang@tum.de} \affiliation{School of Mechanical Engineering,
Shanghai Jiao Tong University, 200240 Shanghai, China}
\affiliation{Institute of Aerodynamics and Fluid Mechanics, Technical
University of Munich, 85748 Garching, Germany}
\author{Huiying Wu}
\email{whysrj@sjtu.edu.cn} \affiliation{School of Mechanical Engineering,
Shanghai Jiao Tong University, 200240 Shanghai, China}
\author{Nikolaus A. Adams}
\email{nikolaus.adams@tum.de} \affiliation{Institute of Aerodynamics and
Fluid Mechanics, Technical University of Munich, 85748 Garching, Germany}
\date{\today}

\begin{abstract}
A novel lattice Boltzmann (LB) model for multiphase flows is developed
that complies with the thermodynamic foundations of kinetic theory. By
directly devising the collision term for LB equation at the discrete
level, a self-tuning equation of state is achieved, which can be
interpreted as the incorporation of short-range molecular interaction. A
pairwise interaction force is introduced to mimic the long-range molecular
interaction, which is responsible for interfacial dynamics. The derived
pressure tensor is naturally consistent with thermodynamic theory, and
surface tension and interface thickness can be independently prescribed.
\end{abstract}

\maketitle

The lattice Boltzmann (LB) method firstly introduced in 1988
\cite{McNamara1988} uses a set of distribution functions with discrete
velocities to depict the complex fluid flows. Due to its kinetic nature, the
LB method shows potential for considering microscopic and mesoscopic
interactions. It is therefore believed that this method is particularly
suitable for multiphase flows, which are complex at the macroscopic level
but are much simpler from the microscopic viewpoint. The applications of LB
method in multiphase flows emerged in the early 1990s \cite{Gunstensen1991}
and have significantly increased in the past decade \cite{Li2016}.

Although various LB models for multiphase flows exist \cite{Shan1993,
Swift1995, He1998}, criticisms have been raised for a long time
\cite{Luo1998, He2002}. In the pseudopotential LB model \cite{Shan1993,
Shan1994}, a pairwise interaction force is used to mimic the microscopic
interaction, which can recover nonideal-gas effects and interfacial dynamics
at the same time. However, such simultaneous recoveries make this model
suffer from thermodynamic inconsistency, though significant progress has
been made in approximating the coexistence densities close to the
thermodynamic results \cite{Sbragaglia2007, Li2012, Khajepor2015}. In the
free-energy LB model \cite{Swift1995, Swift1996}, the thermodynamically
consistent pressure tensor is directly incorporated to produce the dynamics
of multiphase flows. Thus, the annoying evaluations of (high-order)
derivatives are unavoidable, though improvements have been made to remedy
the violation of Galilean invariance in this model \cite{Holdych1998,
Inamuro2000, Kalarakis2002, Kalarakis2003}. Different from the
pseudopotential and free-energy models, the multiphase LB model has also
been developed from kinetic theory via systematic discretization procedures
\cite{He1998, He2002}. Complicated equivalent force terms exist in this
model and severe numerical instability is encountered. Improved models were
formulated \cite{He1999, Lee2005} at the price of sacrificing the underlying
physics and computational simplicity.

In this work, we develop a novel LB model for multiphase flows complying
with the thermodynamic foundations of kinetic theory analyzed by He and
Doolen \cite{He2002}. The underlying molecular interaction responsible for
multiphase flows is divided into the short-range and long-range parts, which
are incorporated by constructing an LB model with self-tuning equation of
state (EOS) and introducing a pairwise interaction force, respectively. The
present LB model has the advantages of the popular pseudopotential and
free-energy LB models and is free of the aforementioned drawbacks.

With the presence of a discrete force term $F_{v,i}$, the LB equation for
the density distribution function $f_i$ can be generally expressed as
\cite{McCracken2005}
\begin{equation} \label{Eq.LBE-V}
\thinmuskip=1.5mu\medmuskip=2mu\thickmuskip=2.5mu
f_i (\mathbf{x} + \mathbf{e}_i \delta_t, t + \delta_t) = f_i +
\delta_t F_{v,i} - {\it \Lambda}_{ik} \big{(} f_k -f_k^\text{eq} +
\tfrac{\delta_t}{2} F_{v,k} \big{)},
\end{equation}
where $\mathbf{e}_i$ is the discrete velocity, ${\it\Lambda}_{ik}$ is the
collision matrix in discrete velocity space, and the right-hand side (RHS),
termed the collision process, is computed at position $\mathbf{x}$ and time
$t$. Owing to the explicit physical significance of the moments of
distribution function, it is more convenient to construct the collision term
in moment space than in discrete velocity space. Orthogonal moments without
weights are adopted \cite{Lallemand2000}, and the RHS of Eq.\
(\ref{Eq.LBE-V}) is transformed into moment space
\begin{equation}\label{Eq.LBE-M}
\bar{\mathbf{m}} = \mathbf{m} + \delta_t \mathbf{F}_m - \mathbf{S} \big{(}
\mathbf{m} - \mathbf{m}^\text{eq} + \tfrac{\delta_t}{2} \mathbf{F}_m
\big{)},
\end{equation}
where $\mathbf{m} = \mathbf{M} (f_i)^\text{T}$ is the rescaled moment with
$\mathbf{M}$ being the dimensionless transformation matrix
\cite{Lallemand2000}, and $\bar{\mathbf{m}}$ denotes the post-collision
moment. For the sake of simplicity, the two-dimensional nine-velocity (D2Q9)
lattice is considered here \cite{Qian1992}, and the extension to
three-dimensional lattice is straightforward though tedious. The equilibrium
moment function $\mathbf{m}^\text{eq} = \mathbf{M} ( f_i^\text{eq}
)^\text{T}$ is devised as
\begin{equation}\label{Eq.meq}
\thinmuskip=1.5mu\medmuskip=2mu\thickmuskip=2.5mu
\begin{split}
\mathbf{m} ^\text{eq} = \big{[} \rho ,\,
2 \alpha_1 \rho + 2 \beta_1 \eta + 3\rho |\hat{\mathbf{u}}|^2 ,\,
\alpha_2 \rho + \beta_2 \eta - 3\rho \times \enskip\quad &\\ |\hat{\mathbf{u}}|^2
+ 9\rho \hat{u}_x^2
\hat{u}_y^2 ,\,
\rho \hat{u}_x ,\,
-\rho \hat{u}_x + 3\rho \hat{u}_x \hat{u}_y^2 ,\,
\rho \hat{u}_y , \quad &\\
-\rho \hat{u}_y + 3\rho \hat{u}_y \hat{u}_x^2 ,\,
\rho (\hat{u}_x^2 - \hat{u}_y^2) ,\,
\rho \hat{u}_x \hat{u}_y \big{]} ^\text{T} & ,
\end{split}
\end{equation}
where $\hat{\mathbf{u}} = \mathbf{u} /c$ with lattice speed $c = \delta_x /
\delta_t$, $\eta$ is introduced to achieve the self-tuning EOS,
$\alpha_{1,2}$ and $\beta_{1,2}$ are coefficients that will be determined
later. The corresponding discrete force term in moment space $\mathbf{F}_m =
\mathbf{M} (F_{v,i}) ^\text{T}$ is set as follows
\begin{equation}\label{Eq.Fm}
\thinmuskip=1.5mu\medmuskip=2mu\thickmuskip=2.5mu
\begin{split}
\mathbf{F}_m = \big{\{} 0 ,\;
6 \hat{\mathbf{F}} \cdot \hat{\mathbf{u}} ,\,
-6 \hat{\mathbf{F}} \cdot \hat{\mathbf{u}} +9 [\hat{\mathbf{F}}
\hat{\mathbf{u}} \hat{\mathbf{u}} \hat{\mathbf{u}}]_{xxyy} ,\,
\hat{F}_x , \qquad\quad & \\
-\hat{F}_x + 3 [\hat{\mathbf{F}} \hat{\mathbf{u}} \hat{\mathbf{u}}]_{xyy} ,\,
\hat{F}_y ,\,
-\hat{F}_y + 3 [\hat{\mathbf{F}} \hat{\mathbf{u}} \hat{\mathbf{u}}]_{xxy} ,
\quad & \\
2 (\hat{F}_x \hat{u}_x - \hat{F}_y \hat{u}_y) ,\,
\hat{F}_x \hat{u}_y + \hat{F}_y \hat{u}_x \big{\}} ^\text{T} &,
\end{split}
\end{equation}
where $\hat{\mathbf{F}} = \mathbf{F} /c$, $[\,\bullet\,]$ denotes
permutation and the subscripts denote tensor indices. In Eqs.\
(\ref{Eq.meq}) and (\ref{Eq.Fm}), the high-order terms of velocity
correspond to the third- and fourth-order Hermite terms in $f_i^\text{eq}$
and $F_{v,i}$, whose effects will be discussed later. The macroscopic
density $\rho$ and velocity $\mathbf{u}$ are defined as
\begin{equation}
\rho = \sum\nolimits_i f_i, \quad
\rho \mathbf{u} = \sum\nolimits_i \mathbf{e}_i f_i + \tfrac{\delta_t}{2}
\mathbf{F}.
\end{equation}

Once the equilibrium distribution function in LB equation is changed to
achieve self-tuning EOS, it has been recognized previously that Newtonian
viscous stress cannot be recovered correctly, and Galilean invariance will
be lost \cite{Swift1996, Dellar2002}. From the Enskog equation for dense
gases in kinetic theory, we note that an extra velocity-dependent term
emerges in the collision term \cite{Chapman1970, Luo1998, He2002}. Inspired
by this fact, some velocity-dependent non-diagonal elements are introduced
in the collision matrix $\mathbf{S}$
\begin{equation}
\setlength{\arraycolsep}{2.5pt}
\mathbf{S} =\! \begin{bmatrix}
s_0^{}& 0& 0& 0& 0& 0& 0& 0& 0\\
0& s_e^{}& k s_\varepsilon \omega_e^{}& 0&
h \hat{u}_x s_q^{} \omega_e& 0&
h \hat{u}_y s_q^{} \omega_e& 0& 0\\
0& 0& s_\varepsilon^{}& 0& 0& 0& 0& 0& 0\\
0& 0& 0& s_j^{}& 0& 0& 0& 0& 0\\
0& 0& 0& 0& s_q^{}& 0& 0& 0& 0\\
0& 0& 0& 0& 0& s_j^{}& 0& 0& 0\\
0& 0& 0& 0& 0& 0& s_q^{}& 0& 0\\
0& 0& 0& 0& 2 b \hat{u}_x s_q^{} \omega_p& 0&
-2 b \hat{u}_y s_q^{} \omega_p& s_p^{}& 0\\
0& 0& 0& 0& b \hat{u}_y s_q^{} \omega_p^{} & 0&
b \hat{u}_x s_q^{} \omega_p^{}& 0& s_p^{}\\
\end{bmatrix},
\end{equation}
where $\omega_{e,p}^{} = s_{e,p}^{}/2 -1$, and $k$, $h$ and $b$ are
coefficients. Note that this improved collision matrix is still invertible.
Through the Chapman-Enskog (CE) analysis and to recover the correct
Newtonian viscous stress (see Supplemental Material for details
\cite{Supplemental}), the coefficients in $\mathbf{m} ^\text{eq}$ and
$\mathbf{S}$ should satisfy
\begin{equation}
\begin{array}{c}
\alpha_2 = - \tfrac{2 \alpha_1 + \varpi +1}{1-\varpi}, \quad
\beta_2 = - \tfrac{2 \beta_1}{1-\varpi}, \\[1.0ex]
k=1-\varpi, \quad
h=\tfrac{6\varpi (1-\varpi)}{1-3\varpi}, \quad
b=\tfrac{1-\varpi}{1-3\varpi},
\end{array}
\end{equation}
where $\varpi$ is related to the bulk viscosity. The recovered macroscopic
equation at the Navier-Stokes level is
\begin{equation}\label{Eq.NSE}
\thinmuskip=1.5mu\medmuskip=2mu\thickmuskip=2.5mu
\begin{cases}
\partial_t \rho + \nabla \cdot (\rho \mathbf{u}) = 0, \\
\partial_t (\rho \mathbf{u}) + \nabla \cdot (\rho \mathbf{uu}) =
-\nabla p_\LBE^{} + \mathbf{F} + \nabla \cdot \mathbf{\Pi} + O(\Ma^3),
\end{cases}
\end{equation}
where $p_\LBE^{}$ is the recovered EOS, $\mathbf{\Pi} = \rho \nu [\nabla
\mathbf{u} + \mathbf{u} \nabla - (\nabla \cdot \mathbf{u}) \mathbf{I} ] +
\rho \varsigma (\nabla \cdot \mathbf{u}) \mathbf{I}$ is the recovered
Newtonian viscous stress, and $\Ma$ denotes the Mach number. Here, the EOS
$p_\LBE^{}$, the kinematic viscosity $\nu$ and the bulk viscosity $\varsigma$
are expressed as
\begin{equation}
\begin{array}{c}
p_\LBE^{} = c_s^2 [(2+\alpha_1) \rho + \beta_1 \eta], \\[1.0ex]
\nu = c_s^2 \big{(} s_p^{-1} - \tfrac12 \big{)} \delta_t, \quad
\varsigma = \varpi c_s^2 \big{(} s_e^{-1} - \tfrac12 \big{)} \delta_t,
\end{array}
\end{equation}
with lattice sound speed $c_s = c / \sqrt{3}$. Obviously, $p_\LBE^{}$ can be
arbitrarily tuned via the built-in $\eta$. Before proceeding further, some
discussion on the present LB model with self-tuning EOS is useful. For the
coefficients $\alpha_{1,2}$ and $\beta_{1,2}$, $\alpha_1 = -1$ and $\beta_1
= 1$ are set as ordinary, and one has $\alpha_2 = 1$. Therefore, $\mathbf{m}
^\text{eq}$ given by Eq.\ (\ref{Eq.meq}) can be decomposed into the ordinary
one derived from the Hermite expansion of the Maxwell-Boltzmann distribution
and the extra one related to $\eta$. The coefficient $\varpi$ plays an
important role here. When $\varpi = 1$, one has $\beta_2 \rightarrow
\infty$, $k = 0$, $h = 0$, and $b = 0$. Thus, $\eta$ should be set to $0$ to
avoid singularity, implying that the present model degenerates into the
classical LB model with ideal-gas EOS. When $\varpi = 1/3$, one has $\beta_2
= -3$, $k = 2/3$, $h \rightarrow \infty$, and $b \rightarrow \infty$. Thus,
the velocity-dependent terms in $\mathbf{S}$ should be removed to avoid
singularity, which means that Newtonian viscous stress cannot be recovered
and Galilean invariance is lost. Compared with previous LB models derived
from the Enskog equation via systematic discretization procedures
\cite{Luo1998, He1998, He2002}, the present model is directly constructed at
the discrete level in moment space and thus is free of complicated
derivative terms, which trigger numerical instability and restrict real
applications of previous models \cite{He1999, Lee2005}.

In the macroscopic equation recovered at the Navier-Stokes level [Eq.\
(\ref{Eq.NSE})], some cubic terms of velocity exist, which are usually
ignored under the low Mach number condition in LB method. By retaining the
high-order terms of velocity in $\mathbf{m} ^\text{eq}$ and $\mathbf{F}_m$
[Eqs.\ (\ref{Eq.meq}) and (\ref{Eq.Fm})], the additional cubic terms can be
partially eliminated \cite{Dellar2014}. To eliminate the remaining cubic
terms, the collision process described by Eq.\ (\ref{Eq.LBE-M}) can be
improved as
\begin{equation}\label{Eq.LBE-M-C}
\begin{split}
\bar{\mathbf{m}} = \mathbf{m} + \delta_t \mathbf{F}_m - \mathbf{S} \left(
\mathbf{m} - \mathbf{m}^\text{eq} + \tfrac{\delta_t}{2} \mathbf{F}_m
\right) & \\
 - \mathbf{R} \left( \mathbf{I} - \tfrac{\mathbf{S}}{2} \right) \left(
\mathbf{m} - \mathbf{m}^\text{eq} + \tfrac{\delta_t}{2} \mathbf{F}_m
\right) & \\
 - \delta_x \mathbf{T} \cdot \nabla \rho - \tfrac{\delta_x}{c^2}
 \mathbf{X} \cdot \nabla p_\LBE^{} &,
\end{split}
\end{equation}
where $\mathbf{R}$, $\mathbf{T}$ and $\mathbf{X}$ are $9 \times 9$, $9
\times 2$ and $9 \times 2$ matrices of order $\Ma^2$, $\Ma^3$ and $\Ma^3$,
respectively, and thus the corresponding terms have negligible effects on
the numerical stability. These correction matrices are set in the following
forms \cite{Huang2018}
\begin{subequations}\label{Eq.CorrectionMatrix}
\newlength{\lsR}\setlength{\lsR}{-0.4ex plus 0ex minus 0ex}
\begin{equation}
\mathbf{R} =
\begin{bmatrix}
0& 0&      0& 0& 0& 0& 0& 0&      0\\[\lsR]
0& R_{11}& 0& 0& 0& 0& 0& R_{17}& R_{18}\\[\lsR]
0& 0&      0& 0& 0& 0& 0& 0&      0\\[\lsR]
0& 0&      0& 0& 0& 0& 0& 0&      0\\[\lsR]
0& 0&      0& 0& 0& 0& 0& 0&      0\\[\lsR]
0& 0&      0& 0& 0& 0& 0& 0&      0\\[\lsR]
0& 0&      0& 0& 0& 0& 0& 0&      0\\[\lsR]
0& R_{71}& 0& 0& 0& 0& 0& R_{77}& R_{78}\\[\lsR]
0& R_{81}& 0& 0& 0& 0& 0& R_{87}& R_{88}\\
\end{bmatrix},
\end{equation}
\begin{equation}
\mathbf{T}  = (\mathbf{0} ,\; \mathbf{T}_1 ,\; \mathbf{0} ,\; \mathbf{0}
,\; \mathbf{0} ,\; \mathbf{0} ,\; \mathbf{0} ,\;
\mathbf{T}_7 ,\; \mathbf{T}_8 ) ^\text{T},
\end{equation}
\begin{equation}
\mathbf{X} = (\mathbf{0} ,\; \mathbf{X}_1 ,\;
\mathbf{0} ,\; \mathbf{0} ,\; \mathbf{0} ,\; \mathbf{0} ,\; \mathbf{0} ,\;
\mathbf{X}_7 ,\; \mathbf{X}_8 ) ^\text{T}.
\end{equation}
\end{subequations}
Through the CE analysis, the nonzero elements in Eq.\
(\ref{Eq.CorrectionMatrix}) can be uniquely and locally determined, and the
truncated term in Eq.\ (\ref{Eq.NSE}) is improved from $O(\Ma^3)$ to
$O(\Ma^5)$ (see Supplemental Material for details \cite{Supplemental}).

As analyzed by He and Doolen \cite{He2002}, a thermodynamically consistent
kinetic model for multiphase flows can be established by combining Enskog
theory for dense gases and mean-field theory for long-range molecular
interaction. In the Enskog equation, short-range molecular interaction is
considered by the collision term, and consequently nonideal-gas EOS is
recovered \cite{Chapman1970}. From this viewpoint, the present LB model with
self-tuning EOS can be interpreted as the incorporation of short-range
molecular interaction, and thus the long-range molecular interaction remains
to be included to construct a valid model for multiphase flows. Following
the idea of the pseudopotential LB model \cite{Shan1993, Shan1994}, a
pairwise interaction force is introduced to mimic the long-range molecular
interaction. Here, nearest-neighbor interaction is considered, and the
interaction force is given as
\begin{equation}\label{Eq.Fint}
\mathbf{F} (\mathbf{x}) = G^2 \rho (\mathbf{x}) \sum\nolimits_i
\omega(|\mathbf{e}_i \delta_t|^2) \rho (\mathbf{x} + \mathbf{e}_i \delta_t)
\mathbf{e}_i \delta_t,
\end{equation}
where $G^2$ is used to control the interaction strength, and the weights
$\omega (\delta_x^2) = 1/3$ and $\omega (2 \delta_x^2) = 1/12$ to maximize
the isotropy degree of $\mathbf{F}$. Note that Eq.\ (\ref{Eq.Fint}) implies
that the long-range molecular interaction is attractive.

The interaction force given by Eq.\ (\ref{Eq.Fint}) is incorporated into the
LB equation via the discrete force term. Based on our previous analysis
\cite{Huang2016}, some $\varepsilon^3 \text{-order}$ terms will be caused by
the discrete lattice effect on the force term, which should be considered
for multiphase flows. In order to cancel such effects, a consistent scheme
for the $\varepsilon^3 \text{-order}$ additional term is employed. The
collision process described by Eq.\ (\ref{Eq.LBE-M-C}) is further improved
as
\begin{equation}\label{Eq.LBE-M-C-3}
\begin{split}
\bar{\mathbf{m}} = \mathbf{m} + \delta_t \mathbf{F}_m - \mathbf{S} \left(
\mathbf{m} - \mathbf{m}^\text{eq} + \tfrac{\delta_t}{2} \mathbf{F}_m
\right) + \mathbf{S} \mathbf{Q}_m & \\
 - \mathbf{R} \left( \mathbf{I} - \tfrac{\mathbf{S}}{2} \right) \left(
\mathbf{m} - \mathbf{m}^\text{eq} + \tfrac{\delta_t}{2} \mathbf{F}_m
\right) & \\
 - \delta_x \mathbf{T} \cdot \nabla \rho - \tfrac{\delta_x}{c^2}
 \mathbf{X} \cdot \nabla p_\LBE^{} &,
\end{split}
\end{equation}
where the discrete additional term $\mathbf{Q}_m$ is set as
\begin{equation}
\mathbf{Q}_m = \Big{(} 0 ,\,
 \tfrac{|\tilde{\mathbf{F}}|^2 }{2} ,\,
-\tfrac{|\tilde{\mathbf{F}}|^2 }{2} ,\,
0 ,\, 0 ,\, 0 ,\, 0 ,\,
\tfrac{\tilde{F}_x^2 - \tilde{F}_y^2}{12},\,
\tfrac{\tilde{F}_x \tilde{F}_y}{12} \Big{)} ^\text{T},
\end{equation}
and $\tilde{\mathbf{F}} = \mathbf{F} / (G \rho c)$. In the CE analysis,
$\mathbf{F}$ is of order $\varepsilon^1$, and thus $\mathbf{Q}_m$ is of
order $\varepsilon^2$. Through the third-order CE analysis, the following
macroscopic equation in steady and stationary situation can be recovered
(see Supplemental Material for details \cite{Supplemental})
\begin{equation}
\thinmuskip=1.5mu\medmuskip=2mu\thickmuskip=2.5mu\begin{cases}
\partial_t \rho = 0, \\
\partial_t (\rho \mathbf{u}) = -\nabla p_\LBE^{} + \mathbf{F} +
\mathbf{R}_\text{iso} + \mathbf{R}_\text{add} + \bar{\mathbf{R}}_\text{iso} +
\bar{\mathbf{R}}_\text{aniso},
\end{cases}
\end{equation}
where $\mathbf{R} _\text{iso} = \tfrac{1}{12} \delta_x^2 \nabla \!\cdot\!
\nabla \mathbf{F}$ is the isotropic term caused by the discrete lattice
effect, $\mathbf{R} _\text{add} = - \tfrac{1}{24} c^2 \nabla \cdot [2
\tilde{\mathbf{F}} \tilde{\mathbf{F}} + (\tilde{\mathbf{F}} \cdot
\tilde{\mathbf{F}}) \mathbf{I} ]$ is the additional term introduced by
$\mathbf{S} \mathbf{Q}_m$ to cancel the effect of $\mathbf{R} _\text{iso}$,
$\bar{ \mathbf{R} } _\text{iso} = -\tfrac{1}{6} \delta_x^2 [(k+1) \tau_e^{}
\tau_q^{} - \tau_p^{} \tau_q^{} ] \nabla \!\cdot\! \nabla \nabla \bar{p}$
and $\bar{\mathbf{R}} _\text{aniso} = - \tfrac{1}{12} \delta_x^2 (12
\tau_p^{} \tau_q^{} - 1) ( \partial_y^2 \partial_x \bar{p}, \, \partial_x^2
\partial_y \bar{p} ) ^\text{T}$ are the isotropic and anisotropic terms
caused by achieving self-tuning EOS, respectively. Here, $\tau _{e, p, q}^{}
= s _{e, p, q} ^{-1} - 1/2$ and $\bar{p} = (3 + \beta_2) c_s^2 \eta$. Note
that $\mathbf{R} _\text{iso}$, $\mathbf{R} _\text{add}$, $\bar{\mathbf{R}}
_\text{iso}$ and $\bar{\mathbf{R}} _\text{aniso}$ are all recovered at the
$\varepsilon^3 \text{-order}$ and thus disappear from the macroscopic
equation at the Navier-Stokes level. For multiphase flows, $\bar{\mathbf{R}}
_\text{iso}$ and $\bar{\mathbf{R}} _\text{aniso}$ should be eliminated by
setting
\begin{equation}
\tau_p^{} \tau_q^{} = (k+1) \tau_e^{} \tau_q^{} = 1/12,
\end{equation}
and $\mathbf{R} _\text{iso}$ and $\mathbf{R} _\text{add}$ can be absorbed
into the pressure tensor. Therefore, the pressure tensor recovered by the
present model is defined as $\nabla \cdot \mathbf{P} = \nabla p_\LBE^{} -
\mathbf{F} - \mathbf{R} _\text{iso} - \mathbf{R} _\text{add} $. By
performing Taylor series expansion of the interaction force, the following
pressure tensor can be derived (see Supplemental Material for details
\cite{Supplemental})
\begin{equation}\label{Eq.PressureTensor}
\mathbf{P} = \big{(} p_\EOS^{} - \kappa \rho \nabla \!\cdot\! \nabla \rho
- \tfrac{\kappa}{2} \nabla \rho \!\cdot\! \nabla \rho \big{)} \mathbf{I}
+ \kappa \nabla \rho \nabla \rho,
\end{equation}
where $\kappa = G^2 \delta_x^4 /4$ and the EOS is
\begin{equation}\label{Eq.recoveredEOS}
p_\EOS^{} = p_\LBE^{} - \tfrac{G^2 \delta_x^2}{2} \rho^2.
\end{equation}
Obviously, $\mathbf{P}$ given by Eq.\ (\ref{Eq.PressureTensor}) is naturally
consistent with thermodynamic theory \cite{Rowlinson1982}, where the free
energy ${\it\Psi}$ is defined as
\begin{equation}
{\it\Psi} = \int_V \big{(} \psi_b + \tfrac{\kappa}{2} |\nabla \rho|^2 \big{)}
d\mathbf{x} .
\end{equation}
Here, $\psi_b$ is the bulk free-energy density related to EOS $p_\EOS^{} = \rho
\partial_\rho \psi_b - \psi_b$, and $\tfrac{\kappa}{2} |\nabla \rho|^2$ is the
interfacial free-energy density. Based on Eq.\ (\ref{Eq.PressureTensor}),
the Maxwell construction can be derived.

In this work, the Carnahan-Starling EOS \cite{Carnahan1969} is taken as an
example
\begin{equation}\label{Eq.CS.EOS}
p_\EOS^{} = K_\EOS \Big{[} \rho R T \tfrac{1 + \vartheta + \vartheta^2 -
\vartheta^3} {(1-\vartheta)^3} - a \rho^2 \Big{]},
\end{equation}
where $R$ is the gas constant, $T$ is the temperature, $\vartheta =
b\rho/4$, $a = 0.496388 R^2 T_c^2 / p_c$ and $b = 0.187295 R T_c / p_c$ with
$T_c$ and $p_c$ denoting the critical temperature and pressure,
respectively. The scaling factor $K _\EOS$ \cite{Wagner2007} is introduced
here to adjust the magnitude of bulk free-energy density $\psi_b$. In the
Carnahan-Starling EOS, the first and second terms describe the effects of
short-range (repulsive) and long-range (attractive) molecular interactions,
respectively \cite{Carnahan1969}. Therefore, a consistency between Eqs.\
(\ref{Eq.recoveredEOS}) and (\ref{Eq.CS.EOS}) can be established, and then
the interaction strength is set as
\begin{equation}
G = K_\INT \sqrt{ 2 K_\EOS a / \delta_x^2} ,
\end{equation}
where the scaling factor $K _\INT$ is introduced to adjust the interfacial
free-energy density $\tfrac{\kappa}{2} |\nabla \rho|^2$, and the lattice
sound speed is chosen as
\begin{equation}
c_s^{} = K_\INT \sqrt{\partial_\rho (p_\EOS^{} + K_\EOS a \rho^2)} \Big{|}
_{ \rho = \rho_l^{} }.
\end{equation}
With this configuration, it is known from thermodynamic theory that the
surface tension $\sigma$ and interface thickness $W$ satisfy
\begin{equation}\label{Eq.W_sigma}
\sigma \propto K_\EOS K_\INT, \quad  W \propto K_\INT,
\end{equation}
which have also been numerically validated (see Supplemental Material for
details \cite{Supplemental}), and where the proportionality constants can be
analytically determined by the pressure tensor. Thus, in real applications
of the present LB model, the surface tension and interface thickness can be
independently prescribed.

\begin{figure}[htbp]
  \centering
  \includegraphics[scale=1,draft=\figdraft]{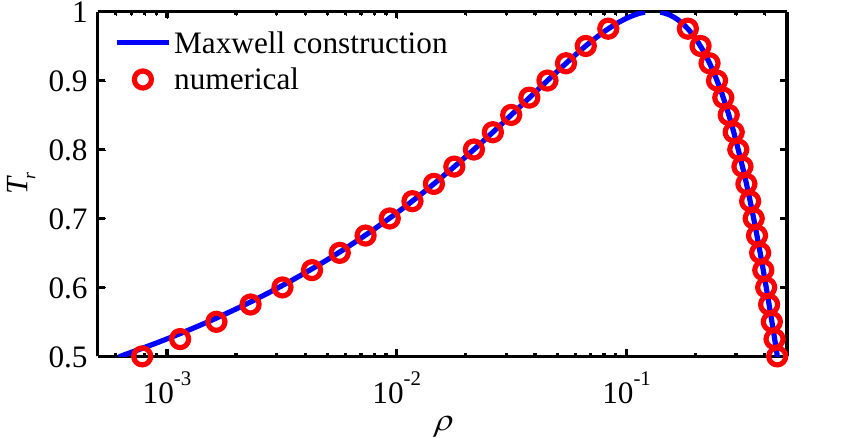}
  \caption{Coexistence curves obtained by simulation of flat interface and
  predicted by Maxwell construction.}
  \label{Fig.01}
\end{figure}

\begin{figure}[htbp]
  \centering
  \includegraphics[scale=1,draft=\figdraft]{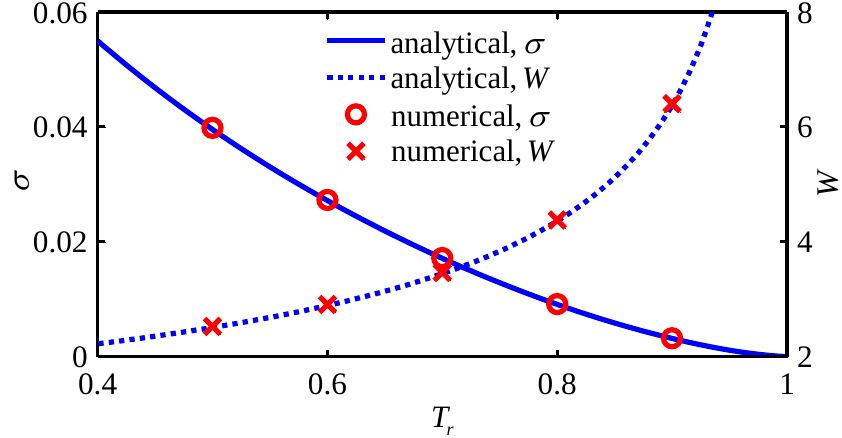}
  \caption{Proportionality constants in Eq.\ (\ref{Eq.W_sigma}) obtained
  by simulation of liquid droplet with various $K_\EOS$ and $K_\INT$ and
  predicted by pressure tensor with $K_\EOS = K_\INT = 1$ for
  flat interface.}
  \label{Fig.02}
\end{figure}

Simulations are performed with $\varpi = 1/6$, $a = 1$, $b = 4$, $R = 1$,
and $\delta_x = 1$, and a detailed implementation of the collision process
[Eq.\ (\ref{Eq.LBE-M-C-3})] is given in Supplemental Material
\cite{Supplemental}. The coexistence curve, as a function of reduced
temperature $T_r$, is firstly computed by simulating a flat interface on a
$1024 \delta_x \times 4 \delta_x$ domain, as shown in Fig.\ \ref{Fig.01}. It
can be seen that the numerical result agrees well with the thermodynamic
result by Maxwell construction. When $T_r < 0.6$, there exists slight
deviation in the gas branch, which is caused by the spatial discretization
error in the interfacial region and can be reduced by increasing the
interface thickness. A liquid droplet is then simulated with various
$K_\EOS$ and $K_\INT$ on a $1024 \delta_x \times 1024 \delta_x$ domain with
the droplet diameter being $512 \delta_x$. Accordingly, the surface tension
is calculated via Laplace's law and the interface thickness is measured from
$\rho = 0.95 \rho_g^{} + 0.05 \rho_l^{}$ to $0.05 \rho_g^{} + 0.95
\rho_l^{}$. Proportionalities described by Eq.\ (\ref{Eq.W_sigma}) can be
accurately observed, and the proportionality constants are in good agreement
with $\sigma$ and $W$ predicted by $\mathbf{P}$ with $K_\EOS = K_\INT = 1$
for flat interface, as shown in Fig.\ \ref{Fig.02}.

\begin{figure}[htbp]
  \centering
  \includegraphics[scale=1,draft=\figdraft]{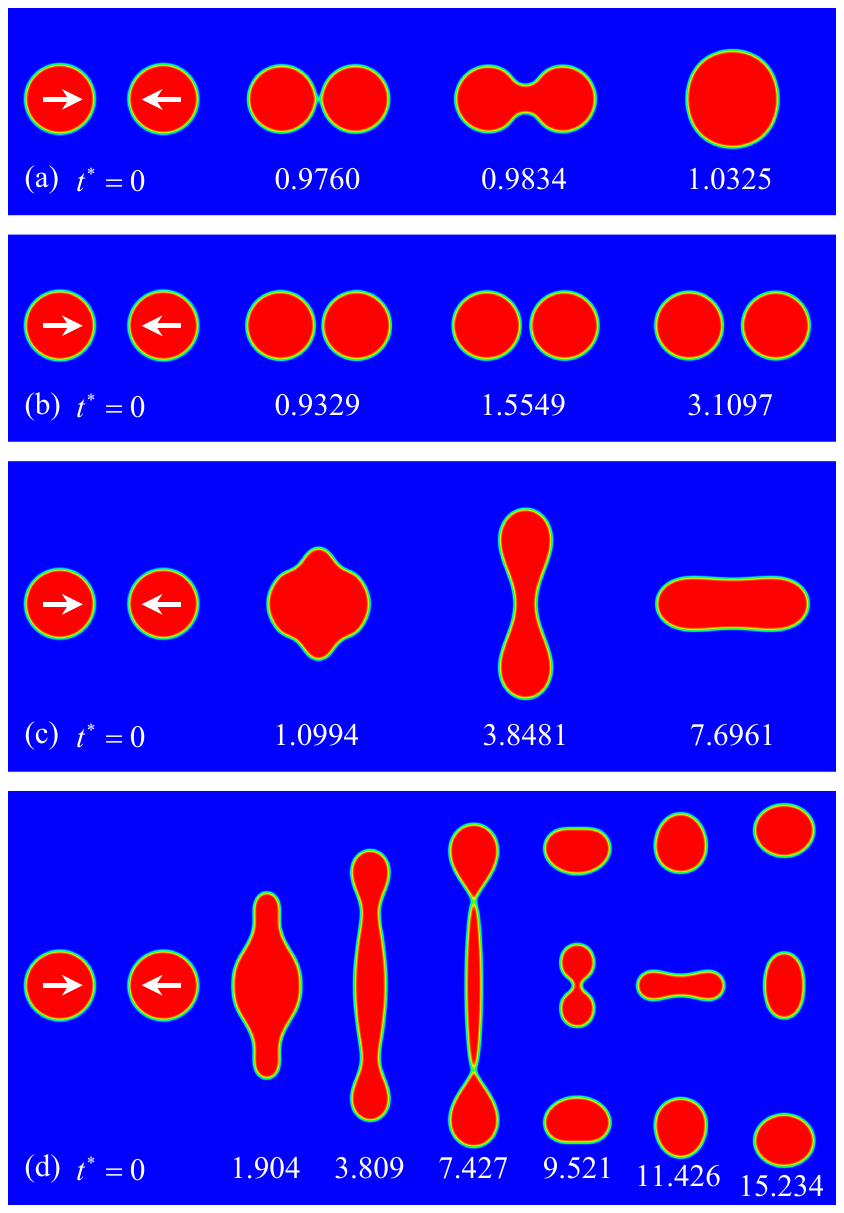}
  \caption{Head-on collision processes of equal-sized droplets at different
  dimensionless time $t^\ast \!= t U/D$ with (a) $\We=0.01$ and $\Rey=1$,
  (b) $\We=0.1$ and $\Rey=1$, (c) $\We=20$ and $\Rey=100$,
  and (d) $\We=60$ and $\Rey=200$.}
  \label{Fig.03}
\end{figure}

As a dynamic problem, oscillation of an elliptic droplet with the
semi-major and minor axes being $96.0 \delta_x$ and $42.7 \delta_x$ is
simulated on a $512 \delta_x \times 512 \delta_x$ domain. Here, $T_r = 0.6$,
$\sigma = 0.01$, and $W = 10$ are chosen. The oscillation period,
numerically measured when the oscillation becomes weak enough, is $18346
\delta_t$, which agrees well with the analytical solution $18628.0 \delta_t$
\cite{Ana}. Head-on collision of equal-sized droplets is further simulated
with $T_r = 0.7$, $\sigma = 0.01$, and $W = 10$. The computational domain
size is $1024 \times 1024$, and the droplet diameter is $128 \delta_x$. The
head-on collision outcome is mainly controlled by the Weber number $\We =
\rho_l^{} U^2 D / \sigma$ and Reynolds number $\Rey = UD / \nu$, with $U$ and
$D$ denoting the relative velocity and droplet diameter, respectively. All
four regimes for head-on collision, experimentally observed by Qian and Law
\cite{Qian1997}, are successfully reproduced here, as shown in Fig.\
\ref{Fig.03}. For $\We = 0.01$ and $\Rey=1$, the droplets approach each other
and then merge with small deformation. As $\We$ increases to $0.1$, the
droplets bounce back without merging. Here, it is interesting to note that
this ``bouncing'' regime has not been observed in previous simulations by
the pseudopotential and free-energy LB models \cite{Lycett-Brown2014,
Moqaddam2016}. For $\We = 20$ and $\Rey=100$, merging happens again,
accompanied with large deformation in this regime. For $\We = 60$ and
$\Rey=200$, the outward motion caused by strong impact splits the merged mass
into three parts, with two main droplets separating from both sides and a
satellite droplet residing at the center, as shown in Fig.\ \ref{Fig.03}(d).

In summary, we have developed a novel LB model for multiphase flows, which
complies with the thermodynamic foundations of kinetic theory and thus is
naturally consistent with thermodynamic theory. The underlying short-range
and long-range molecular interactions are separately incorporated by
constructing an LB model with self-tuning EOS and introducing a pairwise
interaction force. The present model combines the advantages of the popular
pseudopotential and free-energy LB models. Most computations can be carried
out locally, and the surface tension and interface thickness can be
independently prescribed in real applications.

\begin{acknowledgments}
R.H.\ acknowledges the support by the Alexander von Humboldt Foundation,
Germany. This work was also supported by the National Natural Science
Foundation of China through Grant No.\ 51536005.
\end{acknowledgments}

%

\newpage


\section*{Supplementary material}

# Supplemental Material
# Lattice Boltzmann model with self-tuning equation of state for multiphase flows

Rongzong Huang,[1, 2, *] Huiying Wu,[1, †] and Nikolaus A. Adams[2, ‡]

[1]*School of Mechanical Engineering, Shanghai Jiao Tong University, 200240 Shanghai, China*
[2]*Institute of Aerodynamics and Fluid Mechanics,*
*Technical University of Munich, 85748 Garching, Germany*
(Dated: September 7, 2018)

In this Supplemental Material, theoretical analysis of the present LB model for multiphase flows is provided in detail. We start from the collision process given by Eq. (13) in the main text. Both the cubic terms of velocity at the Navier-Stokes level ($\varepsilon^2$-order) and the additional term at the $\varepsilon^3$-order are considered. The collision process, calculated at position $\mathbf{x}$ and time $t$ in moment space, is given as

$$\bar{\mathbf{m}} = \mathbf{m} + \delta_t \mathbf{F}_m - \mathbf{S}\left(\mathbf{m} - \mathbf{m}^{\text{eq}} + \frac{\delta_t}{2}\mathbf{F}_m\right) + \mathbf{S}\mathbf{Q}_m - \mathbf{R}\left(\mathbf{I} - \frac{\mathbf{S}}{2}\right)\left(\mathbf{m} - \mathbf{m}^{\text{eq}} + \frac{\delta_t}{2}\mathbf{F}_m\right) - \delta_x \mathbf{T}\cdot\nabla\rho - \frac{\delta_x}{c^2}\mathbf{X}\cdot\nabla p_{\text{LBE}}, \quad (1)$$

and the streaming process, executed in discrete velocity space, is

$$f_i(\mathbf{x} + \mathbf{e}_i\delta_t, t + \delta_t) = \bar{f}_i(\mathbf{x}, t). \quad (2)$$

The discrete velocity for D2Q9 lattice is [1]

$$\mathbf{e}_i = \begin{cases} c(0, 0)^{\text{T}}, & i = 0, \\ c\big(\cos[(i-1)\pi/2],\ \sin[(i-1)\pi/2]\big)^{\text{T}}, & i = 1, 2, 3, 4, \\ \sqrt{2}c\big(\cos[(2i-1)\pi/4],\ \sin[(2i-1)\pi/4]\big)^{\text{T}}, & i = 5, 6, 7, 8, \end{cases} \quad (3)$$

and the corresponding dimensionless transformation matrix is [2]

$$\mathbf{M} = \begin{bmatrix}
1 & 1 & 1 & 1 & 1 & 1 & 1 & 1 & 1 \\
-4 & -1 & -1 & -1 & -1 & 2 & 2 & 2 & 2 \\
4 & -2 & -2 & -2 & -2 & 1 & 1 & 1 & 1 \\
0 & 1 & 0 & -1 & 0 & 1 & -1 & -1 & 1 \\
0 & -2 & 0 & 2 & 0 & 1 & -1 & -1 & 1 \\
0 & 0 & 1 & 0 & -1 & 1 & 1 & -1 & -1 \\
0 & 0 & -2 & 0 & 2 & 1 & 1 & -1 & -1 \\
0 & 1 & -1 & 1 & -1 & 0 & 0 & 0 & 0 \\
0 & 0 & 0 & 0 & 0 & 1 & -1 & 1 & -1
\end{bmatrix}. \quad (4)$$

## I. CHAPMAN-ENSKOG EXPANSION

Performing Taylor series expansion of $f_i(\mathbf{x} + \mathbf{e}_i\delta_t, t + \delta_t)$ centered at $(\mathbf{x}, t)$ in Eq. (2), and then transforming the result into moment space and combining it with Eq. (1), we can obtain

$$(\mathbf{I}\partial_t + \mathbf{D})\mathbf{m} + \frac{\delta_t}{2}(\mathbf{I}\partial_t + \mathbf{D})^2\mathbf{m} + \frac{\delta_t^2}{6}(\mathbf{I}\partial_t + \mathbf{D})^3\mathbf{m} - \mathbf{F}_m + O(\delta_t^3)$$
$$= -\frac{\mathbf{S}}{\delta_t}\left(\mathbf{m} - \mathbf{m}^{\text{eq}} + \frac{\delta_t}{2}\mathbf{F}_m\right) + \frac{\mathbf{S}}{\delta_t}\mathbf{Q}_m - \frac{\mathbf{R}}{\delta_t}\left(\mathbf{I} - \frac{\mathbf{S}}{2}\right)\left(\mathbf{m} - \mathbf{m}^{\text{eq}} + \frac{\delta_t}{2}\mathbf{F}_m\right) - \frac{\delta_x \mathbf{T}}{\delta_t}\cdot\nabla\rho - \frac{\delta_x \mathbf{X}}{c^2\delta_t}\cdot\nabla p_{\text{LBE}}, \quad (5)$$

where $\mathbf{D} = \mathbf{M}[\text{diag}(\mathbf{e}_i\cdot\nabla)]\mathbf{M}^{-1}$. In this work, the following classical Chapman-Enskog expansions are adopted [3]

$$\partial_t = \sum_{n=1}^{+\infty}\varepsilon^n\partial_{tn},\quad \nabla = \varepsilon^1\nabla_1,\quad f_i = \sum_{n=0}^{+\infty}\varepsilon^n f_i^{(n)},\quad \mathbf{F} = \varepsilon^1\mathbf{F}^{(1)}, \quad (6\text{a})$$

---

[*] rongzong.huang@tum.de
[†] whysrj@sjtu.edu.cn
[‡] nikolaus.adams@tum.de

which indicate

$$\mathbf{D} = \varepsilon^1 \mathbf{D}_1, \quad \mathbf{m} = \sum_{n=0}^{+\infty} \varepsilon^n \mathbf{m}^{(n)}, \quad \mathbf{F}_m = \varepsilon^1 \mathbf{F}_m^{(1)}, \quad \mathbf{Q}_m = \varepsilon^2 \mathbf{Q}_m^{(2)}. \tag{6b}$$

Here, $\varepsilon$ is the small expansion parameter. Substituting the above expansions into Eq. (5), we can rewrite Eq. (5) in the consecutive orders of $\varepsilon$ as follows

$$\varepsilon^0 : -\frac{\mathbf{S}}{\delta_t}(\mathbf{m}^{(0)} - \mathbf{m}^{\text{eq}}) - \frac{\mathbf{R}}{\delta_t}\left(\mathbf{I} - \frac{\mathbf{S}}{2}\right)(\mathbf{m}^{(0)} - \mathbf{m}^{\text{eq}}) = \mathbf{0}, \tag{7a}$$

$$\varepsilon^1 : (\mathbf{I}\partial_{t1} + \mathbf{D}_1)\mathbf{m}^{(0)} - \mathbf{F}_m^{(1)} = -\frac{\mathbf{S}}{\delta_t}\left(\mathbf{m}^{(1)} + \frac{\delta_t}{2}\mathbf{F}_m^{(1)}\right) - \frac{\mathbf{R}}{\delta_t}\left(\mathbf{I} - \frac{\mathbf{S}}{2}\right)\left(\mathbf{m}^{(1)} + \frac{\delta_t}{2}\mathbf{F}_m^{(1)}\right) - \frac{\delta_x \mathbf{T}}{\delta_t} \cdot \nabla_1 \rho - \frac{\delta_x \mathbf{X}}{c^2 \delta_t} \cdot \nabla_1 p_{\text{LBE}}, \tag{7b}$$

$$\varepsilon^2 : \partial_{t2}\mathbf{m}^{(0)} + (\mathbf{I}\partial_{t1} + \mathbf{D}_1)\mathbf{m}^{(1)} + \frac{\delta_t}{2}(\mathbf{I}\partial_{t1} + \mathbf{D}_1)^2 \mathbf{m}^{(0)} = -\frac{\mathbf{S}}{\delta_t}\mathbf{m}^{(2)} + \frac{\mathbf{S}}{\delta_t}\mathbf{Q}_m^{(2)} - \frac{\mathbf{R}}{\delta_t}\left(\mathbf{I} - \frac{\mathbf{S}}{2}\right)\mathbf{m}^{(2)}, \tag{7c}$$

$$\varepsilon^3 : \partial_{t3}\mathbf{m}^{(0)} + \partial_{t2}\mathbf{m}^{(1)} + (\mathbf{I}\partial_{t1} + \mathbf{D}_1)\mathbf{m}^{(2)} + \delta_t(\mathbf{I}\partial_{t1} + \mathbf{D}_1)\partial_{t2}\mathbf{m}^{(0)}$$
$$+ \frac{\delta_t}{2}(\mathbf{I}\partial_{t1} + \mathbf{D}_1)^2 \mathbf{m}^{(1)} + \frac{\delta_t^2}{6}(\mathbf{I}\partial_{t1} + \mathbf{D}_1)^3 \mathbf{m}^{(0)} = -\frac{\mathbf{S}}{\delta_t}\mathbf{m}^{(3)} - \frac{\mathbf{R}}{\delta_t}\left(\mathbf{I} - \frac{\mathbf{S}}{2}\right)\mathbf{m}^{(3)}. \tag{7d}$$

## II. SECOND-ORDER ANALYSIS

Firstly, the standard second-order analysis is carried out to recover the macroscopic equation at the Navier-Stokes level ($\varepsilon^2$-order). To simplify the descriptions, we introduce the following notations

$$\mathbf{G}^{(1)} = \mathbf{m}^{(1)} + \frac{\delta_t}{2}\mathbf{F}_m^{(1)}, \quad \hat{\mathbf{G}}^{(1)} = \left(\mathbf{I} - \frac{\mathbf{S}}{2}\right)\mathbf{G}^{(1)}, \quad \tilde{\mathbf{G}}^{(1)} = \left(\mathbf{I} - \frac{\mathbf{R}}{2}\right)\hat{\mathbf{G}}^{(1)} - \frac{\delta_x \mathbf{T}}{2} \cdot \nabla_1 \rho - \frac{\delta_x \mathbf{X}}{2c^2} \cdot \nabla_1 p_{\text{LBE}}. \tag{8}$$

Therefore, the $\varepsilon^0$-, $\varepsilon^1$-, and $\varepsilon^2$-order equations in Eq. (7) can be reformulated as

$$\varepsilon^0 : -\frac{\mathbf{S}}{\delta_t}(\mathbf{m}^{(0)} - \mathbf{m}^{\text{eq}}) - \frac{\mathbf{R}}{\delta_t}\left(\mathbf{I} - \frac{\mathbf{S}}{2}\right)(\mathbf{m}^{(0)} - \mathbf{m}^{\text{eq}}) = \mathbf{0}, \tag{9a}$$

$$\varepsilon^1 : (\mathbf{I}\partial_{t1} + \mathbf{D}_1)\mathbf{m}^{(0)} - \mathbf{F}_m^{(1)} = -\frac{\mathbf{S}}{\delta_t}\mathbf{G}^{(1)} + \frac{2}{\delta_t}(\tilde{\mathbf{G}}^{(1)} - \hat{\mathbf{G}}^{(1)}), \tag{9b}$$

$$\varepsilon^2 : \partial_{t2}\mathbf{m}^{(0)} + (\mathbf{I}\partial_{t1} + \mathbf{D}_1)\tilde{\mathbf{G}}^{(1)} = -\frac{\mathbf{S}}{\delta_t}\mathbf{m}^{(2)} + \frac{\mathbf{S}}{\delta_t}\mathbf{Q}_m^{(2)} - \frac{\mathbf{R}}{\delta_t}\left(\mathbf{I} - \frac{\mathbf{S}}{2}\right)\mathbf{m}^{(2)}, \tag{9c}$$

where the $\varepsilon^1$-order equation has been used to simplify the $\varepsilon^2$-order equation.

### A. Zeroth-order equation

Based on the zeroth-order ($\varepsilon^0$-order) equation [i.e., Eq. (9a)], we have

$$\varepsilon^0 : \mathbf{m}^{(0)} = \mathbf{m}^{\text{eq}}. \tag{10}$$

Thus, the corresponding equations for conserved moments ($m_0$, $m_3$, and $m_5$) are

$$\varepsilon^0 : \begin{cases} m_0^{(0)} = m_0^{\text{eq}}, \\ m_3^{(0)} = m_3^{\text{eq}}, \\ m_5^{(0)} = m_5^{\text{eq}}. \end{cases} \tag{11}$$

By substituting the Chapman-Enskog expansions [i.e., Eq. (6a)] into the definitions $\rho = \sum_i f_i$ and $\rho \mathbf{u} = \sum_i \mathbf{e}_i f_i + \delta_t \mathbf{F}/2$, and considering Eq. (11), we can obtain

$$\begin{cases} G_0^{(1)} = 0, & m_0^{(n)} = 0 \ (\forall n \geq 2), \\ G_3^{(1)} = 0, & m_3^{(n)} = 0 \ (\forall n \geq 2), \\ G_5^{(1)} = 0, & m_5^{(n)} = 0 \ (\forall n \geq 2), \end{cases} \tag{12}$$

which further indicate that $\tilde{G}_0^{(1)} = \hat{G}_0^{(1)} = 0$, $\tilde{G}_3^{(1)} = \hat{G}_3^{(1)} = 0$, and $\tilde{G}_5^{(1)} = \hat{G}_5^{(1)} = 0$.

### B. First-order equation

To deduce the macroscopic conservation equation, the equations for $m_0$, $m_3$, and $m_5$ are extracted from the first-order ($\varepsilon^1$-order) equation [i.e., Eq. (9b)] as follows

$$\varepsilon^1 : \begin{cases} \partial_{t1} m_0^{(0)} + c\partial_{x1} m_3^{(0)} + c\partial_{y1} m_5^{(0)} - F_{m,0}^{(1)} = -\dfrac{s_0}{\delta_t} G_0^{(1)}, \\ \partial_{t1} m_3^{(0)} + c\partial_{x1}\left(\tfrac{2}{3} m_0^{(0)} + \tfrac{1}{6} m_1^{(0)} + \tfrac{1}{2} m_7^{(0)}\right) + c\partial_{y1} m_8^{(0)} - F_{m,3}^{(1)} = -\dfrac{s_j}{\delta_t} G_3^{(1)}, \\ \partial_{t1} m_5^{(0)} + c\partial_{x1} m_8^{(0)} + c\partial_{y1}\left(\tfrac{2}{3} m_0^{(0)} + \tfrac{1}{6} m_1^{(0)} - \tfrac{1}{2} m_7^{(0)}\right) - F_{m,5}^{(1)} = -\dfrac{s_j}{\delta_t} G_5^{(1)}. \end{cases} \tag{13}$$

Considering Eqs. (10) and (12), Eq. (13) can be finally simplified as

$$\varepsilon^1 : \begin{cases} \partial_{t1} \rho + \nabla_1 \cdot (\rho \mathbf{u}) = 0, \\ \partial_{t1} (\rho \mathbf{u}) + \nabla_1 \cdot (\rho \mathbf{uu}) = -\nabla_1 p_{\text{LBE}} + \mathbf{F}^{(1)}, \end{cases} \tag{14}$$

where the recovered EOS $p_{\text{LBE}}$ is

$$p_{\text{LBE}} = c_s^2 [(2 + \alpha_1)\rho + \beta_1 \eta], \tag{15}$$

and $c_s = c/\sqrt{3}$ is the lattice sound speed. Based on Eq. (14), the following relations for time derivative terms can be derived

$$\begin{aligned} \partial_{t1} \rho &= -\nabla_1 \cdot (\rho \mathbf{u}), \\ \partial_{t1} (\rho \mathbf{u}) &= -\nabla_1 p_{\text{LBE}} + \mathbf{F}^{(1)} - \nabla_1 \cdot (\rho \mathbf{uu}), \\ \partial_{t1} (\rho \mathbf{uu}) &= -[(\nabla_1 p_{\text{LBE}})\mathbf{u}] + [\mathbf{F}^{(1)} \mathbf{u}] - \nabla_1 \cdot (\rho \mathbf{uuu}), \\ \partial_{t1} (\rho \mathbf{uuu}) &= -[(\nabla_1 p_{\text{LBE}})\mathbf{uu}] + [\mathbf{F}^{(1)} \mathbf{uu}] - \nabla_1 \cdot (\rho \mathbf{uuuu}), \\ \partial_{t1} (\rho \mathbf{uuuu}) &= -[(\nabla_1 p_{\text{LBE}})\mathbf{uuu}] + [\mathbf{F}^{(1)} \mathbf{uuu}] - \nabla_1 \cdot (\rho \mathbf{uuuuu}), \end{aligned} \tag{16}$$

in which $[\bullet]$ denotes permutation, e.g., $[(\nabla_1 p_{\text{LBE}})\mathbf{uu}] = (\nabla_1 p_{\text{LBE}})\mathbf{uu} + \mathbf{u}(\nabla_1 p_{\text{LBE}})\mathbf{u} + \mathbf{uu}(\nabla_1 p_{\text{LBE}})$ and $[\mathbf{F}^{(1)} \mathbf{uu}] = \mathbf{F}^{(1)} \mathbf{uu} + \mathbf{u} \mathbf{F}^{(1)} \mathbf{u} + \mathbf{uu} \mathbf{F}^{(1)}$. Here, we would like to point out that the last term $\nabla_1 \cdot (\rho \mathbf{uuu})$ in the relation for $\partial_{t1}(\rho \mathbf{uu})$ corresponds to the cubic terms of velocity in the recovered macroscopic equation at Navier-Stokes level for the classical LB model, which are usually ignored under the low Mach number condition.

### C. Second-order equation

Extracting the equations for $m_0$, $m_3$, and $m_5$ from the second-order ($\varepsilon^2$-order) equation [i.e., Eq. (9c)], we have

$$\varepsilon^2 : \begin{cases} \partial_{t2} m_0^{(0)} + \partial_{t1} \tilde{G}_0^{(1)} + c\partial_{x1} \tilde{G}_3^{(1)} + c\partial_{y1} \tilde{G}_5^{(1)} = -\dfrac{s_0}{\delta_t} m_0^{(2)}, \\ \partial_{t2} m_3^{(0)} + \partial_{t1} \tilde{G}_3^{(1)} + c\partial_{x1}\left(\tfrac{2}{3} \tilde{G}_0^{(1)} + \tfrac{1}{6} \tilde{G}_1^{(1)} + \tfrac{1}{2} \tilde{G}_7^{(1)}\right) + c\partial_{y1} \tilde{G}_8^{(1)} = -\dfrac{s_j}{\delta_t} m_3^{(2)}, \\ \partial_{t2} m_5^{(0)} + \partial_{t1} \tilde{G}_5^{(1)} + c\partial_{x1} \tilde{G}_8^{(1)} + c\partial_{y1}\left(\tfrac{2}{3} \tilde{G}_0^{(1)} + \tfrac{1}{6} \tilde{G}_1^{(1)} - \tfrac{1}{2} \tilde{G}_7^{(1)}\right) = -\dfrac{s_j}{\delta_t} m_5^{(2)}. \end{cases} \tag{17}$$

With the help of Eqs. (10) and (12), Eq. (17) can be simplified as

$$\varepsilon^2 : \begin{cases} \partial_{t2}\rho = 0, \\ \partial_{t2}(\rho\mathbf{u}) = \nabla_1 \cdot \mathbf{\Pi}^{(1)}, \end{cases} \tag{18}$$

where the viscous stress tensor $\mathbf{\Pi}^{(1)}$ is given as

$$\mathbf{\Pi}^{(1)} = -c^2 \begin{bmatrix} \frac{1}{2}\tilde{G}_7^{(1)} & \tilde{G}_8^{(1)} \\ \tilde{G}_8^{(1)} & -\frac{1}{2}\tilde{G}_7^{(1)} \end{bmatrix} - c^2 \begin{bmatrix} \frac{1}{6}\tilde{G}_1^{(1)} & 0 \\ 0 & \frac{1}{6}\tilde{G}_1^{(1)} \end{bmatrix}. \tag{19}$$

*1. Newtonian viscous stress*

In order to calculate the viscous stress tensor given by Eq. (19), the equations for the related moments ($m_1$, $m_7$, and $m_8$) at the order of $\varepsilon^1$ are extracted from Eq. (9b) as follows

$$\begin{aligned} -\frac{1}{\delta_t}\big(s_e G_1^{(1)} + k s_\varepsilon \omega_e G_2^{(1)} + h\hat{u}_x s_q \omega_e G_4^{(1)} + h\hat{u}_y s_q \omega_e G_6^{(1)}\big) + \frac{2}{\delta_t}\big(\tilde{G}_1^{(1)} - \hat{G}_1^{(1)}\big) \\ = \partial_{t1} m_1^{(0)} + c\partial_{x1}\big(m_3^{(0)} + m_4^{(0)}\big) + c\partial_{y1}\big(m_5^{(0)} + m_6^{(0)}\big) - F_{m,1}^{(1)}, \end{aligned} \tag{20a}$$

$$\begin{aligned} -\frac{1}{\delta_t}\big(s_p G_7^{(1)} + 2b\hat{u}_x s_q \omega_p G_4^{(1)} - 2b\hat{u}_y s_q \omega_p G_6^{(1)}\big) + \frac{2}{\delta_t}\big(\tilde{G}_7^{(1)} - \hat{G}_7^{(1)}\big) \\ = \partial_{t1} m_7^{(0)} + c\partial_{x1}\big(\tfrac{1}{3}m_3^{(0)} - \tfrac{1}{3}m_4^{(0)}\big) - c\partial_{y1}\big(\tfrac{1}{3}m_5^{(0)} - \tfrac{1}{3}m_6^{(0)}\big) - F_{m,7}^{(1)}, \end{aligned} \tag{20b}$$

$$\begin{aligned} -\frac{1}{\delta_t}\big(s_p G_8^{(1)} + b\hat{u}_y s_q \omega_p G_4^{(1)} + b\hat{u}_x s_q \omega_p G_6^{(1)}\big) + \frac{2}{\delta_t}\big(\tilde{G}_8^{(1)} - \hat{G}_8^{(1)}\big) \\ = \partial_{t1} m_8^{(0)} + c\partial_{x1}\big(\tfrac{2}{3}m_5^{(0)} + \tfrac{1}{3}m_6^{(0)}\big) + c\partial_{y1}\big(\tfrac{2}{3}m_3^{(0)} + \tfrac{1}{3}m_4^{(0)}\big) - F_{m,8}^{(1)}, \end{aligned} \tag{20c}$$

where the involved $\varepsilon^1$-order terms $G_2^{(1)}$, $G_4^{(1)}$, and $G_6^{(1)}$ can also be obtained from Eq. (9b)

$$-\frac{1}{\delta_t} s_\varepsilon G_2^{(1)} = \partial_{t1} m_2^{(0)} + c\partial_{x1} m_4^{(0)} + c\partial_{y1} m_6^{(0)} - F_{m,2}^{(1)}, \tag{21a}$$

$$-\frac{1}{\delta_t} s_q G_4^{(1)} = \partial_{t1} m_4^{(0)} + c\partial_{x1}\big(\tfrac{1}{3}m_1^{(0)} + \tfrac{1}{3}m_2^{(0)} - m_7^{(0)}\big) + c\partial_{y1} m_8^{(0)} - F_{m,4}^{(1)}, \tag{21b}$$

$$-\frac{1}{\delta_t} s_q G_6^{(1)} = \partial_{t1} m_6^{(0)} + c\partial_{x1} m_8^{(0)} + c\partial_{y1}\big(\tfrac{1}{3}m_1^{(0)} + \tfrac{1}{3}m_2^{(0)} + m_7^{(0)}\big) - F_{m,6}^{(1)}. \tag{21c}$$

Multiplying Eqs. (21a), (21b), and (21c) by $k$, $h\hat{u}_x$, and $h\hat{u}_y$, respectively, and then adding the results to Eq. (20a), we can obtain

$$\begin{aligned} -\frac{1}{\delta_t}\frac{2s_e}{2-s_e}\hat{G}_1^{(1)} + \frac{2}{\delta_t}\big(\tilde{G}_1^{(1)} - \hat{G}_1^{(1)}\big) = &\, \partial_{t1} m_1^{(0)} + c\partial_{x1}\big(m_3^{(0)} + m_4^{(0)}\big) + c\partial_{y1}\big(m_5^{(0)} + m_6^{(0)}\big) - F_{m,1}^{(1)} \\ &+ k\big[\partial_{t1} m_2^{(0)} + c\partial_{x1} m_4^{(0)} + c\partial_{y1} m_6^{(0)} - F_{m,2}^{(1)}\big] \\ &+ h\hat{u}_x\big[\partial_{t1} m_4^{(0)} + c\partial_{x1}\big(\tfrac{1}{3}m_1^{(0)} + \tfrac{1}{3}m_2^{(0)} - m_7^{(0)}\big) + c\partial_{y1} m_8^{(0)} - F_{m,4}^{(1)}\big] \\ &+ h\hat{u}_y\big[\partial_{t1} m_6^{(0)} + c\partial_{x1} m_8^{(0)} + c\partial_{y1}\big(\tfrac{1}{3}m_1^{(0)} + \tfrac{1}{3}m_2^{(0)} + m_7^{(0)}\big) - F_{m,6}^{(1)}\big]. \end{aligned} \tag{22}$$

With the help of Eqs. (10) and (16), as well as $\mathbf{m}^{\mathrm{eq}}$ and $\mathbf{F}_m$, Eq. (22) can be simplified after lengthy algebra as follows

$$\begin{aligned} -\frac{1}{\delta_t}\frac{2s_e}{2-s_e}\hat{G}_1^{(1)} + \frac{2}{\delta_t}\big(\tilde{G}_1^{(1)} - \hat{G}_1^{(1)}\big) = &-(2\alpha_1 + k\alpha_2 + k)\nabla_1\cdot(\rho\mathbf{u}) + \tfrac{h[(\alpha_2-4)\beta_1 - (\alpha_1+2)\beta_2]}{3\beta_1}\mathbf{u}\cdot\nabla_1\rho \\ &+ \tfrac{h(3\beta_1+\beta_2)-6\beta_1(1-k)}{\beta_1}\mathbf{u}\cdot\nabla_1\hat{p}_{\mathrm{LBE}} + (2\beta_1 + k\beta_2)\partial_{t1}\eta \\ &- c\big\{9(2k+h)\hat{u}_x\hat{u}_y^2\partial_{x1}\hat{p}_{\mathrm{LBE}} + 9(2k+h)\hat{u}_x^2\hat{u}_y\partial_{y1}\hat{p}_{\mathrm{LBE}}\big\} \\ &- c\big\{3[(1-k)\hat{u}_x^3 - (2k+h)\hat{u}_x\hat{u}_y^2]\partial_{x1}\rho + 3[(1-k)\hat{u}_y^3 - (2k+h)\hat{u}_x^2\hat{u}_y]\partial_{y1}\rho\big\} \\ &- c\left\{\begin{aligned}&[9(1-k)\hat{u}_x^2 - 2(3k+h)\hat{u}_y^2]\rho\partial_{x1}\hat{u}_x + [9(1-k)\hat{u}_y^2 - 2(3k+h)\hat{u}_x^2]\rho\partial_{y1}\hat{u}_y \\ &\qquad - 4(3k+h)\hat{u}_x\hat{u}_y\rho(\partial_{x1}\hat{u}_y + \partial_{y1}\hat{u}_x)\end{aligned}\right\} \\ &- c\big\{9k\nabla_1\cdot(\rho\hat{\mathbf{u}}\hat{u}_x^2\hat{u}_y^2) + 3h\hat{u}_y\partial_{x1}(\rho\hat{u}_x^3\hat{u}_y) + 3h\hat{u}_x\partial_{y1}(\rho\hat{u}_x\hat{u}_y^3)\big\}, \end{aligned} \tag{23}$$

where $\hat{p}_{\text{LBE}} = p_{\text{LBE}}/c^2$. In order to correctly recover the Newtonian viscous stress, one should have

$$\begin{cases} 2\alpha_1 + k\alpha_2 + k = \dfrac{h[(\alpha_2 - 4)\beta_1 - (\alpha_1 + 2)\beta_2]}{3\beta_1} \equiv -2\varpi, \\ \dfrac{h(3\beta_1 + \beta_2) - 6\beta_1(1-k)}{\beta_1} = 0, \\ 2\beta_1 + k\beta_2 = 0. \end{cases} \quad (24)$$

Here, a crucial parameter $\varpi$ is further introduced, and it is worth emphasizing that $\varpi$ cannot be simply set to 1 following the classical LB model with ideal-gas EOS. Otherwise, singularity will be encountered [see Eq. (31)] and the present model should degenerate into the classical LB model. Moreover, there exist some third- and fifth-order terms of velocity in Eq. (23), which can be simply ignored under the low Mach number condition. As a better choice, eliminating the third-order terms will be discussed in next subsection.

Multiplying Eqs. (21b) and (21c) by $2b\hat{u}_x$ and $-2b\hat{u}_y$, respectively, and then adding the results to Eq. (20b), we can obtain

$$\begin{aligned} -\frac{1}{\delta_t}\frac{2s_p}{2-s_p}\hat{G}_7^{(1)} + \frac{2}{\delta_t}\big(\tilde{G}_7^{(1)} - \hat{G}_7^{(1)}\big) =\ & \partial_{t1}m_7^{(0)} + c\partial_{x1}\big(\tfrac{1}{3}m_3^{(0)} - \tfrac{1}{3}m_4^{(0)}\big) - c\partial_{y1}\big(\tfrac{1}{3}m_5^{(0)} - \tfrac{1}{3}m_6^{(0)}\big) - F_{m,7}^{(1)} \\ & + 2b\hat{u}_x\big[\partial_{t1}m_4^{(0)} + c\partial_{x1}\big(\tfrac{1}{3}m_1^{(0)} + \tfrac{1}{3}m_2^{(0)} - m_7^{(0)}\big) + c\partial_{y1}m_8^{(0)} - F_{m,4}^{(1)}\big] \\ & - 2b\hat{u}_y\big[\partial_{t1}m_6^{(0)} + c\partial_{x1}m_8^{(0)} + c\partial_{y1}\big(\tfrac{1}{3}m_1^{(0)} + \tfrac{1}{3}m_2^{(0)} + m_7^{(0)}\big) - F_{m,6}^{(1)}\big]. \end{aligned} \quad (25)$$

By utilizing Eqs. (10) and (16), as well as $\mathbf{m}^{\text{eq}}$ and $\mathbf{F}_m$, Eq. (25) can be simplified after lengthy algebra as follows

$$\begin{aligned} -\frac{1}{\delta_t}\frac{2s_p}{2-s_p}\hat{G}_7^{(1)} + \frac{2}{\delta_t}\big(\tilde{G}_7^{(1)} - \hat{G}_7^{(1)}\big) =\ & \tfrac{2}{3}[\partial_{x1}(\rho u_x) - \partial_{y1}(\rho u_y)] + \tfrac{2}{3}\tfrac{b[(\alpha_2-4)\beta_1-(\alpha_1+2)\beta_2]}{\beta_1}(u_x\partial_{x1}\rho - u_y\partial_{y1}\rho) \\ & + \tfrac{(6b-2)\beta_1 + 2b\beta_2}{\beta_1}(u_x\partial_{x1}\hat{p}_{\text{LBE}} - u_y\partial_{y1}\hat{p}_{\text{LBE}}) \\ & + c\big\{6b\hat{u}_x\hat{u}_y^2\partial_{x1}\hat{p}_{\text{LBE}} - 6b\hat{u}_x^2\hat{u}_y\partial_{y1}\hat{p}_{\text{LBE}}\big\} \\ & - c\big\{(\hat{u}_x^3 + 2b\hat{u}_x\hat{u}_y^2)\partial_{x1}\rho - (\hat{u}_y^3 + 2b\hat{u}_x^2\hat{u}_y)\partial_{y1}\rho\big\} \\ & - c\big\{(3\hat{u}_x^2 + 4b\hat{u}_y^2)\rho\partial_{x1}\hat{u}_x - (3\hat{u}_y^2 + 4b\hat{u}_x^2)\rho\partial_{y1}\hat{u}_y\big\} \\ & + c\big\{6b\hat{u}_y\partial_{x1}(\rho\hat{u}_x^3\hat{u}_y) - 6b\hat{u}_x\partial_{y1}(\rho\hat{u}_x\hat{u}_y^3)\big\}. \end{aligned} \quad (26)$$

In order to correctly recover the Newtonian viscous stress, one should have

$$\begin{cases} \dfrac{b[(\alpha_2 - 4)\beta_1 - (\alpha_1 + 2)\beta_2]}{\beta_1} = -1, \\ \dfrac{(6b-2)\beta_1 + 2b\beta_2}{\beta_1} = 0. \end{cases} \quad (27)$$

Multiplying Eqs. (21b) and (21c) by $b\hat{u}_y$ and $b\hat{u}_x$, respectively, and then adding the results to Eq. (20c), we can obtain

$$\begin{aligned} -\frac{1}{\delta_t}\frac{2s_p}{2-s_p}\hat{G}_8^{(1)} + \frac{2}{\delta_t}\big(\tilde{G}_8^{(1)} - \hat{G}_8^{(1)}\big) =\ & \partial_{t1}m_8^{(0)} + c\partial_{x1}\big(\tfrac{2}{3}m_5^{(0)} + \tfrac{1}{3}m_6^{(0)}\big) + c\partial_{y1}\big(\tfrac{2}{3}m_3^{(0)} + \tfrac{1}{3}m_4^{(0)}\big) - F_{m,8}^{(1)} \\ & + b\hat{u}_y\big[\partial_{t1}m_4^{(0)} + c\partial_{x1}\big(\tfrac{1}{3}m_1^{(0)} + \tfrac{1}{3}m_2^{(0)} - m_7^{(0)}\big) + c\partial_{y1}m_8^{(0)} - F_{m,4}^{(1)}\big] \\ & + b\hat{u}_x\big[\partial_{t1}m_6^{(0)} + c\partial_{x1}m_8^{(0)} + c\partial_{y1}\big(\tfrac{1}{3}m_1^{(0)} + \tfrac{1}{3}m_2^{(0)} + m_7^{(0)}\big) - F_{m,6}^{(1)}\big]. \end{aligned} \quad (28)$$

With the help of Eqs. (10) and (16), as well as $\mathbf{m}^{\text{eq}}$ and $\mathbf{F}_m$, Eq. (28) can be simplified as follows

$$\begin{aligned} -\frac{1}{\delta_t}\frac{2s_p}{2-s_p}\hat{G}_8^{(1)} + \frac{2}{\delta_t}\big(\tilde{G}_8^{(1)} - \hat{G}_8^{(1)}\big) =\ & \tfrac{1}{3}[\partial_{x1}(\rho u_y) + \partial_{y1}(\rho u_x)] + \tfrac{1}{3}\tfrac{b[(\alpha_2-4)\beta_1-(\alpha_1+2)\beta_2]}{\beta_1}(u_y\partial_{x1}\rho + u_x\partial_{y1}\rho) \\ & + \tfrac{(3b-1)\beta_1 + b\beta_2}{\beta_1}(u_y\partial_{x1}\hat{p}_{\text{LBE}} + u_x\partial_{y1}\hat{p}_{\text{LBE}}) \\ & - c\big\{3b(\hat{u}_y^3 + 2\hat{u}_x^2\hat{u}_y)\partial_{x1}\hat{p}_{\text{LBE}} + 3b(\hat{u}_x^3 + 2\hat{u}_x\hat{u}_y^2)\partial_{y1}\hat{p}_{\text{LBE}}\big\} \\ & + c\big\{b(\hat{u}_y^3 + 2\hat{u}_x^2\hat{u}_y)\partial_{x1}\rho + b(\hat{u}_x^3 + 2\hat{u}_x\hat{u}_y^2)\partial_{y1}\rho\big\} \\ & + c\big\{2b\hat{u}_x\hat{u}_y\rho\partial_{x1}\hat{u}_x + 2b\hat{u}_x\hat{u}_y\rho\partial_{y1}\hat{u}_y + 2b(\hat{u}_x^2 + \hat{u}_y^2)\rho(\partial_{x1}\hat{u}_y + \partial_{y1}\hat{u}_x)\big\} \\ & - c\big\{3b\hat{u}_x\partial_{x1}(\rho\hat{u}_x^3\hat{u}_y) + 3b\hat{u}_y\partial_{y1}(\rho\hat{u}_x\hat{u}_y^3)\big\}. \end{aligned} \quad (29)$$

Similarly, in order to correctly recover the Newtonian viscous stress, one should have

$$\begin{cases} \dfrac{b[(\alpha_2 - 4)\beta_1 - (\alpha_1 + 2)\beta_2]}{\beta_1} = -1, \\ \dfrac{(3b-1)\beta_1 + b\beta_2}{\beta_1} = 0. \end{cases} \quad (30)$$

Here, we note that Eq. (30) is actually the same as Eq. (27), which indicates that the diagonal and non-diagonal elements in the traceless part of viscous stress tensor can be correctly recovered, or not, at the same time.

Eqs. (24), (27), and (30) give the constraints on the coefficients in $\mathbf{m}^{\text{eq}}$ and $\mathbf{S}$ required for the correct recovery of Newtonian viscous stress. Considering the classical LB model can be viewed as a special case of the present LB model, Eqs. (24), (27), and (30) must be compatible with each other. Solving these equations, we have

$$\alpha_2 = -\frac{2\alpha_1 + \varpi + 1}{1 - \varpi}, \quad \beta_2 = -\frac{2\beta_1}{1 - \varpi}, \quad k = 1 - \varpi, \quad h = \frac{6\varpi(1-\varpi)}{1 - 3\varpi}, \quad b = \frac{1 - \varpi}{1 - 3\varpi}. \quad (31)$$

In practical applications, $\alpha_1$ is set to $-1$ according to the ordinary equilibrium moment function derived from the Maxwell-Boltzmann distribution, and $\beta_1$ is set to 1 as ordinary. Therefore, all the coefficients $\alpha_2$, $\beta_2$, $k$, $h$, and $b$ are uniquely determined by $\varpi$.

### 2. Cubic terms of velocity

From Eqs. (23), (26), and (29), and considering $\mathbf{R} \sim O(Ma^2)$, $\mathbf{T} \sim O(Ma^3)$, and $\mathbf{X} \sim O(Ma^3)$, we have $\hat{G}_1^{(1)} \sim O(Ma)$, $\hat{G}_7^{(1)} \sim O(Ma)$, and $\hat{G}_8^{(1)} \sim O(Ma)$. Thus, the leading-order terms of Eqs. (23), (26), and (29) are given as

$$\begin{cases} -\dfrac{1}{\delta_t}\dfrac{2s_e}{2-s_e}\hat{G}_1^{(1)} = 2\varpi\rho(\partial_{x1}u_x + \partial_{y1}u_y) + O(Ma^3), \\ -\dfrac{1}{\delta_t}\dfrac{2s_p}{2-s_p}\hat{G}_7^{(1)} = \tfrac{2}{3}\rho(\partial_{x1}u_x - \partial_{y1}u_y) + O(Ma^3), \\ -\dfrac{1}{\delta_t}\dfrac{2s_p}{2-s_p}\hat{G}_8^{(1)} = \tfrac{1}{3}\rho(\partial_{x1}u_y + \partial_{y1}u_x) + O(Ma^3). \end{cases} \quad (32)$$

Based on Eq. (23) and $\tilde{G}_1^{(1)} = \hat{G}_1^{(1)} - \tfrac{1}{2}\bigl(R_{11}\hat{G}_1^{(1)} + R_{17}\hat{G}_7^{(1)} + R_{18}\hat{G}_8^{(1)} + \delta_x\mathbf{T}_1\cdot\nabla_1\rho + \delta_x\mathbf{X}_1\cdot\nabla_1\hat{p}_{\text{LBE}}\bigr)$, we have

$$\begin{aligned}
-\frac{1}{\delta_t}\frac{2s_e}{2-s_e}\tilde{G}_1^{(1)} =\ & 2\varpi\rho(\partial_{x1}u_x+\partial_{y1}u_y) + \tfrac{1}{\delta_t}\tfrac{2}{2-s_e}\bigl(R_{11}\hat{G}_1^{(1)}+R_{17}\hat{G}_7^{(1)}+R_{18}\hat{G}_8^{(1)}+\delta_x\mathbf{T}_1\cdot\nabla_1\rho+\delta_x\mathbf{X}_1\cdot\nabla_1\hat{p}_{\text{LBE}}\bigr) \\
& - c\bigl\{9(2k+h)\hat{u}_x\hat{u}_y^2\partial_{x1}\hat{p}_{\text{LBE}} + 9(2k+h)\hat{u}_x^2\hat{u}_y\partial_{y1}\hat{p}_{\text{LBE}}\bigr\} \\
& - c\bigl\{3[(1-k)\hat{u}_x^3 - (2k+h)\hat{u}_x\hat{u}_y^2]\partial_{x1}\rho + 3[(1-k)\hat{u}_y^3 - (2k+h)\hat{u}_x^2\hat{u}_y]\partial_{y1}\rho\bigr\} \\
& - c\begin{Bmatrix} [9(1-k)\hat{u}_x^2 - 2(3k+h)\hat{u}_y^2]\rho\partial_{x1}\hat{u}_x + [9(1-k)\hat{u}_y^2 - 2(3k+h)\hat{u}_x^2]\rho\partial_{y1}\hat{u}_y \\ -4(3k+h)\hat{u}_x\hat{u}_y\rho(\partial_{x1}\hat{u}_y + \partial_{y1}\hat{u}_x) \end{Bmatrix} \\
& - c\bigl\{9k\nabla_1\cdot(\rho\hat{\mathbf{u}}\hat{u}_x^2\hat{u}_y^2) + 3h\hat{u}_y\partial_{x1}(\rho\hat{u}_x^3\hat{u}_y) + 3h\hat{u}_x\partial_{y1}(\rho\hat{u}_x\hat{u}_y^3)\bigr\}.
\end{aligned} \quad (33)$$

In order to eliminate the cubic terms of velocity in Eq. (33) and with the consideration of Eq. (32), we can set

$$R_{11} = -\frac{s_e}{4\varpi}[9(1-k) - 2(3k+h)](\hat{u}_x^2 + \hat{u}_y^2), \quad R_{17} = -\frac{3s_p(2-s_e)}{4(2-s_p)}[9(1-k) + 2(3k+h)](\hat{u}_x^2 - \hat{u}_y^2),$$

$$R_{18} = \frac{12s_p(2-s_e)}{2-s_p}(3k+h)\hat{u}_x\hat{u}_y, \quad \mathbf{T}_1 = \frac{3(2-s_e)}{2}\begin{bmatrix}(1-k)\hat{u}_x^3 - (2k+h)\hat{u}_x\hat{u}_y^2 \\ (1-k)\hat{u}_y^3 - (2k+h)\hat{u}_x^2\hat{u}_y\end{bmatrix}, \quad \mathbf{X}_1 = \frac{9(2-s_e)}{2}(2k+h)\begin{bmatrix}\hat{u}_x\hat{u}_y^2 \\ \hat{u}_x^2\hat{u}_y\end{bmatrix}. \quad (34)$$

Thus, Eq. (33) can be finally simplified as

$$-\frac{1}{\delta_t}\frac{2s_e}{2-s_e}\tilde{G}_1^{(1)} = 2\varpi\rho(\partial_{x1}u_x + \partial_{y1}u_y) + O(Ma^5). \quad (35)$$

From Eq. (26), and considering $\tilde{G}_7^{(1)} = \hat{G}_7^{(1)} - \frac{1}{2}\big(R_{71}\hat{G}_1^{(1)} + R_{77}\hat{G}_7^{(1)} + R_{78}\hat{G}_8^{(1)} + \delta_x \mathbf{T}_7 \cdot \nabla_1 \rho + \delta_x \mathbf{X}_7 \cdot \nabla_1 \hat{p}_{\text{LBE}}\big)$, we have

$$\begin{aligned}
-\frac{1}{\delta_t}\frac{2s_p}{2-s_p}\tilde{G}_7^{(1)} =& \tfrac{2}{3}\rho(\partial_{x1}u_x - \partial_{y1}u_y) + \tfrac{1}{\delta_t}\tfrac{2}{2-s_p}\big(R_{71}\hat{G}_1^{(1)} + R_{77}\hat{G}_7^{(1)} + R_{78}\hat{G}_8^{(1)} + \delta_x \mathbf{T}_7 \cdot \nabla_1 \rho + \delta_x \mathbf{X}_7 \cdot \nabla_1 \hat{p}_{\text{LBE}}\big) \\
& + c\big\{6b\hat{u}_x\hat{u}_y^2\partial_{x1}\hat{p}_{\text{LBE}} - 6b\hat{u}_x^2\hat{u}_y\partial_{y1}\hat{p}_{\text{LBE}}\big\} - c\big\{(\hat{u}_x^3 + 2b\hat{u}_x\hat{u}_y^2)\partial_{x1}\rho - (\hat{u}_y^3 + 2b\hat{u}_x^2\hat{u}_y)\partial_{y1}\rho\big\} \\
& - c\big\{(3\hat{u}_x^2 + 4b\hat{u}_y^2)\rho\partial_{x1}\hat{u}_x - (3\hat{u}_y^2 + 4b\hat{u}_x^2)\rho\partial_{y1}\hat{u}_y\big\} + c\big\{6b\hat{u}_y\partial_{x1}(\rho\hat{u}_x^3\hat{u}_y) - 6b\hat{u}_x\partial_{y1}(\rho\hat{u}_x\hat{u}_y^3)\big\}.
\end{aligned} \quad (36)$$

To eliminate the additional cubic terms of velocity, and with the consideration of Eq. (32), we can set

$$\begin{aligned}
R_{71} &= -\frac{s_e(2-s_p)}{4\varpi(2-s_e)}(3-4b)(\hat{u}_x^2 - \hat{u}_y^2), \quad R_{77} = -\frac{3s_p}{4}(3+4b)(\hat{u}_x^2 + \hat{u}_y^2), \\
R_{78} &= 0, \quad \mathbf{T}_7 = \frac{2-s_p}{2}\begin{bmatrix} \hat{u}_x^3 + 2b\hat{u}_x\hat{u}_y^2 \\ -\hat{u}_y^3 - 2b\hat{u}_x^2\hat{u}_y \end{bmatrix}, \quad \mathbf{X}_7 = -3(2-s_p)b\begin{bmatrix} \hat{u}_x\hat{u}_y^2 \\ -\hat{u}_x^2\hat{u}_y \end{bmatrix}.
\end{aligned} \quad (37)$$

Thus, Eq. (36) is finally simplified as

$$-\frac{1}{\delta_t}\frac{2s_p}{2-s_p}\tilde{G}_7^{(1)} = \tfrac{2}{3}\rho(\partial_{x1}u_x - \partial_{y1}u_y) + O(Ma^5). \quad (38)$$

Based on Eq. (29), and considering $\tilde{G}_8^{(1)} = \hat{G}_8^{(1)} - \frac{1}{2}\big(R_{81}\hat{G}_1^{(1)} + R_{87}\hat{G}_7^{(1)} + R_{88}\hat{G}_8^{(1)} + \delta_x \mathbf{T}_8 \cdot \nabla_1 \rho + \delta_x \mathbf{X}_8 \cdot \nabla_1 \hat{p}_{\text{LBE}}\big)$, we have

$$\begin{aligned}
-\frac{1}{\delta_t}\frac{2s_p}{2-s_p}\tilde{G}_8^{(1)} =& \tfrac{1}{3}\rho(\partial_{x1}u_y + \partial_{y1}u_x) + \tfrac{1}{\delta_t}\tfrac{2}{2-s_p}\big(R_{81}\hat{G}_1^{(1)} + R_{87}\hat{G}_7^{(1)} + R_{88}\hat{G}_8^{(1)} + \delta_x \mathbf{T}_8 \cdot \nabla_1 \rho + \delta_x \mathbf{X}_8 \cdot \nabla_1 \hat{p}_{\text{LBE}}\big) \\
& - c\big\{3b(\hat{u}_y^3 + 2\hat{u}_x^2\hat{u}_y)\partial_{x1}\hat{p}_{\text{LBE}} + 3b(\hat{u}_x^3 + 2\hat{u}_x\hat{u}_y^2)\partial_{y1}\hat{p}_{\text{LBE}}\big\} \\
& + c\big\{b(\hat{u}_y^3 + 2\hat{u}_x^2\hat{u}_y)\partial_{x1}\rho + b(\hat{u}_x^3 + 2\hat{u}_x\hat{u}_y^2)\partial_{y1}\rho\big\} \\
& + c\big\{2b\hat{u}_x\hat{u}_y\rho\partial_{x1}\hat{u}_x + 2b\hat{u}_x\hat{u}_y\rho\partial_{y1}\hat{u}_y + 2b(\hat{u}_x^2 + \hat{u}_y^2)\rho(\partial_{x1}\hat{u}_y + \partial_{y1}\hat{u}_x)\big\} \\
& - c\big\{3b\hat{u}_x\partial_{x1}(\rho\hat{u}_x^3\hat{u}_y) + 3b\hat{u}_y\partial_{y1}(\rho\hat{u}_x\hat{u}_y^3)\big\}.
\end{aligned} \quad (39)$$

Similarly, to eliminate the additional cubic terms, we can set

$$\begin{aligned}
R_{81} &= \frac{s_e(2-s_p)}{\varpi(2-s_e)}b\hat{u}_x\hat{u}_y, \quad R_{87} = 0, \\
R_{88} &= 6s_p b(\hat{u}_x^2 + \hat{u}_y^2), \quad \mathbf{T}_8 = -\frac{2-s_p}{2}b\begin{bmatrix} \hat{u}_y^3 + 2\hat{u}_x^2\hat{u}_y \\ \hat{u}_x^3 + 2\hat{u}_x\hat{u}_y^2 \end{bmatrix}, \quad \mathbf{X}_8 = \frac{3(2-s_p)}{2}b\begin{bmatrix} \hat{u}_y^3 + 2\hat{u}_x^2\hat{u}_y \\ \hat{u}_x^3 + 2\hat{u}_x\hat{u}_y^2 \end{bmatrix},
\end{aligned} \quad (40)$$

and then Eq. (39) is simplified as

$$-\frac{1}{\delta_t}\frac{2s_p}{2-s_p}\tilde{G}_8^{(1)} = \tfrac{1}{3}\rho(\partial_{x1}u_y + \partial_{y1}u_x) + O(Ma^5). \quad (41)$$

As given by Eqs. (34), (37), and (40), the nonzero elements in $\mathbf{R}$, $\mathbf{T}$, and $\mathbf{X}$ can be uniquely and locally determined, and the results are also consistent with the aforementioned conditions $\mathbf{R} \sim O(Ma^2)$, $\mathbf{T} \sim O(Ma^3)$, and $\mathbf{X} \sim O(Ma^3)$. Based on Eqs. (35), (38), and (41), the viscous stress tensor given by Eq. (19) can be simplified as

$$\mathbf{\Pi}^{(1)} = \rho\nu[\nabla_1\mathbf{u} + \mathbf{u}\nabla_1 - (\nabla_1 \cdot \mathbf{u})\mathbf{I}] + \rho\varsigma(\nabla_1 \cdot \mathbf{u})\mathbf{I} + O(Ma^5), \quad (42)$$

where the kinematic viscosity $\nu$ and bulk viscosity $\varsigma$ are given as

$$\nu = c_s^2\big(s_p^{-1} - \tfrac{1}{2}\big)\delta_t, \qquad \varsigma = \varpi c_s^2\big(s_e^{-1} - \tfrac{1}{2}\big)\delta_t. \quad (43)$$

### D. Macroscopic equation

Combining the $\varepsilon^1$- and $\varepsilon^2$-order equations [i.e., Eqs. (14) and (18)] and utilizing Eq. (42), the macroscopic equation at the Navier-Stokes level can be finally recovered as follows

$$\begin{cases} \partial_t \rho + \nabla \cdot (\rho\mathbf{u}) = 0, \\ \partial_t(\rho\mathbf{u}) + \nabla \cdot (\rho\mathbf{u}\mathbf{u}) = -\nabla p_{\text{LBE}} + \mathbf{F} + \nabla \cdot \{\rho\nu[\nabla\mathbf{u} + \mathbf{u}\nabla - (\nabla \cdot \mathbf{u})\mathbf{I}] + \rho\varsigma(\nabla \cdot \mathbf{u})\mathbf{I}\} + O(Ma^5). \end{cases} \quad (44)$$

Obviously, the Newtonian viscous stress is correctly recovered and thus the Galilean invariance can be satisfied. From the above second-order analysis, we can see that the force term $\mathbf{F}$ can be recovered with no discrete lattice effect at the $\varepsilon^2$-order, and the term $\mathbf{SQ}_m$ in LB equation for $\varepsilon^3$-order additional term makes no difference to the recovered macroscopic equation at the $\varepsilon^2$-order.

## III. THIRD-ORDER ANALYSIS

The target of the present third-order analysis is to identify the leading-order terms at the $\varepsilon^3$-order, which should be seriously considered for multiphase flows [4]. These leading-order terms are mainly related to the density gradient in the interfacial region caused by the pairwise interaction force, which means that such terms are irrelevant to time and velocity. Therefore, a steady and stationary situation can be considered, which is simple but effective, and can also avoid leading to the Burnett level equation that is unnecessary and undesirable. In a steady and stationary situation, the $\varepsilon^0$-, $\varepsilon^1$-, $\varepsilon^2$-, and $\varepsilon^3$-order equations in Eq. (7) can be simplified as

$$\varepsilon^0 : \mathbf{m}^{(0)} = \mathbf{m}^{\mathrm{eq}}, \tag{45a}$$

$$\varepsilon^1 : \partial_{t1}\mathbf{m}^{(0)} + \mathbf{D}_1 \mathbf{m}^{(0)} - \mathbf{F}_m^{(1)} = -\frac{\mathbf{S}}{\delta_t}\left(\mathbf{m}^{(1)} + \frac{\delta_t}{2}\mathbf{F}_m^{(1)}\right), \tag{45b}$$

$$\varepsilon^2 : \partial_{t2}\mathbf{m}^{(0)} - \delta_t \mathbf{D}_1 \left(\mathbf{S}^{-1} - \frac{\mathbf{I}}{2}\right)\left(\mathbf{D}_1\mathbf{m}^{(0)} - \mathbf{F}_m^{(1)}\right) = -\frac{\mathbf{S}}{\delta_t}\mathbf{m}^{(2)} + \frac{\mathbf{S}}{\delta_t}\mathbf{Q}_m^{(2)}, \tag{45c}$$

$$\varepsilon^3 : \partial_{t3}\mathbf{m}^{(0)} + \delta_t^2\left[\mathbf{D}_1\left(\mathbf{S}^{-1} - \frac{\mathbf{I}}{2}\right)\mathbf{D}_1\left(\mathbf{S}^{-1} - \frac{\mathbf{I}}{2}\right)\left(\mathbf{D}_1\mathbf{m}^{(0)} - \mathbf{F}_m^{(1)}\right) - \frac{1}{12}\mathbf{D}_1^3\mathbf{m}^{(0)}\right] + \mathbf{D}_1\mathbf{Q}_m^{(2)} = -\frac{\mathbf{S}}{\delta_t}\mathbf{m}^{(3)}, \tag{45d}$$

where the terms $\partial_{t1}\mathbf{m}^{(0)}$, $\partial_{t2}\mathbf{m}^{(0)}$, and $\partial_{t3}\mathbf{m}^{(0)}$ are reserved as a gauge for the equations at different orders. Here, it is worth pointing out that the present improved collision matrix $\mathbf{S}$ is invertible, and the inverse matrix is given as

$$\mathbf{S}^{-1} = \begin{bmatrix} s_0^{-1} & 0 & 0 & 0 & 0 & 0 & 0 & 0 & 0 \\ 0 & s_e^{-1} & k(s_e^{-1} - \frac{1}{2}) & 0 & h\hat{u}_x(s_e^{-1} - \frac{1}{2}) & 0 & h\hat{u}_y(s_e^{-1} - \frac{1}{2}) & 0 & 0 \\ 0 & 0 & s_\varepsilon^{-1} & 0 & 0 & 0 & 0 & 0 & 0 \\ 0 & 0 & 0 & s_j^{-1} & 0 & 0 & 0 & 0 & 0 \\ 0 & 0 & 0 & 0 & s_q^{-1} & 0 & 0 & 0 & 0 \\ 0 & 0 & 0 & 0 & 0 & s_j^{-1} & 0 & 0 & 0 \\ 0 & 0 & 0 & 0 & 0 & 0 & s_q^{-1} & 0 & 0 \\ 0 & 0 & 0 & 0 & 2b\hat{u}_x(s_p^{-1} - \frac{1}{2}) & 0 & -2b\hat{u}_y(s_p^{-1} - \frac{1}{2}) & s_p^{-1} & 0 \\ 0 & 0 & 0 & 0 & b\hat{u}_y(s_p^{-1} - \frac{1}{2}) & 0 & b\hat{u}_x(s_p^{-1} - \frac{1}{2}) & 0 & s_p^{-1} \end{bmatrix}. \tag{46}$$

Similar to the second-order analysis, the equations for $m_0$, $m_3$, and $m_5$ are extracted from Eq. (45) to deduce the macroscopic conservation equation. From the $\varepsilon^0$-order equation [i.e., Eq. (45a)], we have

$$\varepsilon^0 : \begin{cases} m_0^{(0)} = m_0^{\mathrm{eq}}, \\ m_3^{(0)} = m_3^{\mathrm{eq}}, \\ m_5^{(0)} = m_5^{\mathrm{eq}}, \end{cases} \tag{47a}$$

which indicate

$$\begin{cases} m_0^{(1)} + \frac{\delta_t}{2}F_{m,0}^{(1)} = 0, & m_0^{(n)} = 0 \ (\forall n \geq 2), \\ m_3^{(1)} + \frac{\delta_t}{2}F_{m,3}^{(1)} = 0, & m_3^{(n)} = 0 \ (\forall n \geq 2), \\ m_5^{(1)} + \frac{\delta_t}{2}F_{m,5}^{(1)} = 0, & m_5^{(n)} = 0 \ (\forall n \geq 2), \end{cases} \tag{47b}$$

by considering the definition equations for $\rho$ and $\rho\mathbf{u}$. The equations for conserved moments in the $\varepsilon^1$-order equation [i.e., Eq. (45b)] can be simplified as

$$\varepsilon^1 : \begin{cases} \partial_{t1}\rho = 0, \\ \partial_{t1}(\rho u_x) = -\partial_{x1}p_{\mathrm{LBE}} + F_x^{(1)}, \\ \partial_{t1}(\rho u_y) = -\partial_{y1}p_{\mathrm{LBE}} + F_y^{(1)}, \end{cases} \tag{48}$$

where the recovered EOS $p_{\text{LBE}} = c_s^2[(2+\alpha_1)\rho + \beta_1\eta]$ and the lattice sound speed $c_s = c/\sqrt{3}$. Similarly, the equations for conserved moments in the $\varepsilon^2$-order equation [i.e., Eq. (45c)] can be simplified as

$$\varepsilon^2 : \begin{cases} \partial_{t2}\rho = 0, \\ \partial_{t2}(\rho u_x) = 0, \\ \partial_{t2}(\rho u_y) = 0. \end{cases} \quad (49)$$

After some lengthy algebra, the equations for conserved moments in the $\varepsilon^3$-order equation [i.e., Eq. (45d)] can be finally simplified as

$$\varepsilon^3 : \begin{cases} \partial_{t3}\rho = 0, \\ \partial_{t3}(\rho u_x) = \frac{1}{12}\delta_x^2 \nabla_1 \cdot \nabla_1 F_x^{(1)} - \frac{1}{24}c^2[2\nabla_1\cdot(\tilde{\mathbf{F}}^{(1)}\tilde{F}_x^{(1)}) + \partial_{x1}(\tilde{\mathbf{F}}^{(1)}\cdot\tilde{\mathbf{F}}^{(1)})] \\ \qquad\qquad - \frac{1}{6}\delta_x^2[(k+1)\tau_e\tau_q - \tau_p\tau_q]\nabla_1\cdot\nabla_1\partial_{x1}\bar{p} - \frac{1}{12}\delta_x^2(12\tau_p\tau_q - 1)\partial_{y1}^2\partial_{x1}\bar{p}, \\ \partial_{t3}(\rho u_y) = \frac{1}{12}\delta_x^2 \nabla_1 \cdot \nabla_1 F_y^{(1)} - \frac{1}{24}c^2[2\nabla_1\cdot(\tilde{\mathbf{F}}^{(1)}\tilde{F}_y^{(1)}) + \partial_{y1}(\tilde{\mathbf{F}}^{(1)}\cdot\tilde{\mathbf{F}}^{(1)})] \\ \qquad\qquad - \frac{1}{6}\delta_x^2[(k+1)\tau_e\tau_q - \tau_p\tau_q]\nabla_1\cdot\nabla_1\partial_{y1}\bar{p} - \frac{1}{12}\delta_x^2(12\tau_p\tau_q - 1)\partial_{x1}^2\partial_{y1}\bar{p}, \end{cases} \quad (50)$$

where $\tau_{e,p,q} = s_{e,p,q}^{-1} - 1/2$ and $\bar{p} = c_s^2[(3\alpha_1 + \alpha_2 + 2)\rho + (3\beta_1 + \beta_2)\eta]$. With the ordinary choices $\alpha_1 = -1$ and $\beta_1 = 1$, $\bar{p}$ is then given as $\bar{p} = (3+\beta_2)c_s^2\eta$. Combining the $\varepsilon^1$-, $\varepsilon^2$-, and $\varepsilon^3$-order equations [i.e., Eqs. (48), (49), and (50)], the macroscopic equation in steady and stationary situation at the $\varepsilon^3$-order can be recovered as follows

$$\begin{cases} \partial_t\rho = 0, \\ \partial_t(\rho\mathbf{u}) = -\nabla p_{\text{LBE}} + \mathbf{F} + \mathbf{R}_{\text{iso}} + \mathbf{R}_{\text{add}} + \bar{\mathbf{R}}_{\text{iso}} + \bar{\mathbf{R}}_{\text{aniso}}, \end{cases} \quad (51)$$

where $\mathbf{R}_{\text{iso}}$ denotes the isotropic term caused by the discrete lattice effect on the force term, $\mathbf{R}_{\text{add}}$ denotes the additional term introduced by the term $\mathbf{SQ}_m$ in LB equation, $\bar{\mathbf{R}}_{\text{iso}}$ and $\bar{\mathbf{R}}_{\text{aniso}}$ denote the isotropic and anisotropic terms, respectively, caused by $\eta$ that is introduced to achieve self-tuning EOS, and these $\varepsilon^3$-order terms are expressed as

$$\begin{aligned} \mathbf{R}_{\text{iso}} &= \frac{1}{12}\delta_x^2 \nabla\cdot\nabla\mathbf{F}, & \mathbf{R}_{\text{add}} &= -\frac{1}{24}c^2\nabla\cdot[2\tilde{\mathbf{F}}\tilde{\mathbf{F}} + (\tilde{\mathbf{F}}\cdot\tilde{\mathbf{F}})\mathbf{I}], \\ \bar{\mathbf{R}}_{\text{iso}} &= -\frac{1}{6}\delta_x^2[(k+1)\tau_e\tau_q - \tau_p\tau_q]\nabla\cdot\nabla\nabla\bar{p}, & \bar{\mathbf{R}}_{\text{aniso}} &= -\frac{1}{12}\delta_x^2(12\tau_p\tau_q - 1)\big(\partial_y^2\partial_x\bar{p},\ \partial_x^2\partial_y\bar{p}\big)^{\text{T}}. \end{aligned} \quad (52)$$

For multiphase flows, the anisotropic term $\bar{\mathbf{R}}_{\text{aniso}}$ must be eliminated based on our previous analysis [4], and the isotropic term $\bar{\mathbf{R}}_{\text{iso}}$ should also be eliminated since it is related to the relaxation parameters. For these purposes, we can set

$$\tau_p\tau_q = (k+1)\tau_e\tau_q = \frac{1}{12}. \quad (53)$$

### A. Pressure tensor

With the consideration of the $\varepsilon^3$-order terms $\mathbf{R}_{\text{iso}}$ and $\mathbf{R}_{\text{add}}$, the pressure tensor $\mathbf{P}$ recovered by the present LB model is defined as

$$\nabla \cdot \mathbf{P} = \nabla p_{\text{LBE}} - \mathbf{F} - \mathbf{R}_{\text{iso}} - \mathbf{R}_{\text{add}}. \quad (54)$$

Performing Taylor series expansion of $\rho(\mathbf{x} + \mathbf{e}_i\delta_t)$ centered at $\mathbf{x}$, the interaction force $\mathbf{F}$ can be expressed as

$$\mathbf{F} = G^2\delta_x^2\rho\nabla\rho + \frac{G^2\delta_x^4}{6}\rho\nabla\nabla\cdot\nabla\rho + O(\partial_{x,y}^5). \quad (55)$$

Therefore, $\mathbf{F} + \mathbf{R}_{\text{iso}}$ in Eq. (54) can be simplified as

$$\begin{aligned} \mathbf{F} + \mathbf{R}_{\text{iso}} &= G^2\delta_x^2\rho\nabla\rho + \frac{G^2\delta_x^4}{6}\rho\nabla\nabla\cdot\nabla\rho + \frac{G^2\delta_x^4}{12}\nabla\cdot\nabla(\rho\nabla\rho) + O(\partial_{x,y}^5) \\ &= \frac{G^2\delta_x^2}{2}\nabla\cdot(\rho^2\mathbf{I}) + \frac{G^2\delta_x^4}{6}\nabla\cdot[a_1\nabla\rho\nabla\rho + a_2\rho\nabla\nabla\rho + (a_3\nabla\rho\cdot\nabla\rho + a_4\rho\nabla\cdot\nabla\rho)\mathbf{I}] \\ &\quad + \frac{G^2\delta_x^4}{12}\nabla\cdot[b_1\nabla\rho\nabla\rho + b_1\rho\nabla\nabla\rho + (b_2\nabla\rho\cdot\nabla\rho + b_2\rho\nabla\cdot\nabla\rho)\mathbf{I}] + O(\partial_{x,y}^5), \end{aligned} \quad (56)$$

where $a_{1,2,3,4}$ and $b_{1,2}$ are free parameters that satisfy

$$a_1 + a_2 + 2a_3 = 0, \qquad a_1 + a_4 = 0, \qquad a_2 + a_4 = 1; \qquad b_1 + b_2 = 1. \tag{57}$$

Although these free parameters cannot be completely determined, they make no difference to the thermodynamic properties predicted by the pressure tensor [4]. Therefore, we choose $2a_2 + b_1 = 0$ as a supplementary constraint for the sake of simplicity. Then, Eq. (56) can be further simplified as

$$\mathbf{F} + \mathbf{R}_{\mathrm{iso}} = \frac{G^2\delta_x^2}{2}\nabla\cdot(\rho^2\mathbf{I}) + \frac{G^2\delta_x^4}{12}\nabla\cdot[-2\nabla\rho\nabla\rho + (2\nabla\rho\cdot\nabla\rho + 3\rho\nabla\cdot\nabla\rho)\mathbf{I}] + O(\partial_{x,y}^5). \tag{58}$$

By utilizing Eq. (55), $\mathbf{R}_{\mathrm{add}}$ can be simplified as

$$\mathbf{R}_{\mathrm{add}} = -\frac{G^2\delta_x^4}{24}\nabla\cdot[2\nabla\rho\nabla\rho + (\nabla\rho\cdot\nabla\rho)\mathbf{I}] + O(\partial_{x,y}^5). \tag{59}$$

Based on Eqs. (58) and (59), the pressure tensor defined by Eq. (54) can be finally derived as follows

$$\begin{aligned}\mathbf{P} &= \left(p_{\mathrm{LBE}} - \frac{G^2\delta_x^2}{2}\rho^2 - \frac{G^2\delta_x^4}{4}\rho\nabla\cdot\nabla\rho - \frac{G^2\delta_x^4}{8}\nabla\rho\cdot\nabla\rho\right)\mathbf{I} + \frac{G^2\delta_x^4}{4}\nabla\rho\nabla\rho + O(\partial_{x,y}^4) \\ &= \left(p_{\mathrm{EOS}} - \kappa\rho\nabla\cdot\nabla\rho - \frac{\kappa}{2}\nabla\rho\cdot\nabla\rho\right)\mathbf{I} + \kappa\nabla\rho\nabla\rho + O(\partial_{x,y}^4),\end{aligned} \tag{60}$$

where $G^2\delta_x^4/4$ is denoted by $\kappa$, and $p_{\mathrm{LBE}} - G^2\delta_x^2\rho^2/2$ is, namely, the EOS $p_{\mathrm{EOS}}$.

## IV. IMPLEMENTATION

At first glance, the present collision process [Eq. (1)] seems difficult to implement. However, in real applications, it can be executed in the following sequence

(1) $\begin{cases} \bar{\mathbf{m}} \leftarrow \mathbf{m}, \\ \mathbf{m} \leftarrow \mathbf{m} - \mathbf{m}^{\mathrm{eq}}, \\ \bar{\mathbf{m}} \leftarrow \bar{\mathbf{m}} - 2\mathbf{m}, \\ \mathbf{m} \leftarrow \mathbf{m} + \frac{\delta_t}{2}\mathbf{F}_m; \end{cases}$

(2) $\begin{cases} m_1 \leftarrow m_1 + \frac{1}{2}ks_\varepsilon m_2 + \frac{1}{2}hs_q(\hat{u}_x m_4 + \hat{u}_y m_6), \\ m_7 \leftarrow m_7 + bs_q(\hat{u}_x m_4 - \hat{u}_y m_6), \\ m_8 \leftarrow m_8 + \frac{1}{2}bs_q(\hat{u}_y m_4 + \hat{u}_x m_6); \end{cases}$

(3) $\begin{cases} \mathbf{m} \leftarrow [\mathbf{I} - \frac{1}{2}\mathrm{diag}(\mathbf{S})]\mathbf{m}; \end{cases}$

(4) $\begin{cases} \bar{m}_1 \leftarrow \bar{m}_1 - R_{11}m_1 - R_{17}m_7 - R_{18}m_8 - \delta_x\mathbf{T}_1\cdot\nabla\rho - \frac{\delta_x}{c^2}\mathbf{X}_1\cdot\nabla p_{\mathrm{LBE}}, \\ \bar{m}_7 \leftarrow \bar{m}_7 - R_{71}m_1 - R_{77}m_7 - R_{78}m_8 - \delta_x\mathbf{T}_7\cdot\nabla\rho - \frac{\delta_x}{c^2}\mathbf{X}_7\cdot\nabla p_{\mathrm{LBE}}, \\ \bar{m}_8 \leftarrow \bar{m}_8 - R_{81}m_1 - R_{87}m_7 - R_{88}m_8 - \delta_x\mathbf{T}_8\cdot\nabla\rho - \frac{\delta_x}{c^2}\mathbf{X}_8\cdot\nabla p_{\mathrm{LBE}}; \end{cases}$

(5) $\begin{cases} \bar{\mathbf{m}} \leftarrow \bar{\mathbf{m}} + 2\mathbf{m}; \end{cases}$

(6) $\begin{cases} \bar{m}_1 \leftarrow \bar{m}_1 + s_e Q_{m,1} + k(\frac{1}{2}s_e - 1)s_\varepsilon Q_{m,2}, \\ \bar{m}_2 \leftarrow \bar{m}_2 + s_\varepsilon Q_{m,2}, \\ \bar{m}_7 \leftarrow \bar{m}_7 + s_p Q_{m,7}, \\ \bar{m}_8 \leftarrow \bar{m}_8 + s_p Q_{m,8}; \end{cases}$

where "$\leftarrow$" indicates assignment, and $\mathrm{diag}(\mathbf{S})$ denotes the diagonal part of the present collision matrix $\mathbf{S}$. Here, steps (1), (3), and (5) are the same as those for the classical multiple-relaxation-time (MRT) collision process, step (2) corresponds to the improvement of the collision matrix, step (4) corresponds to the elimination of the cubic terms of velocity, and step (6) corresponds to the introduction of the $\varepsilon^3$-order additional term. In step (4), $\nabla\rho$ and $\nabla p_{\mathrm{LBE}}$ can be locally evaluated resorting to the interaction force as follows

$$\begin{aligned}\nabla\rho &= \frac{1}{\delta_x^2}\sum_i \omega(|\mathbf{e}_i\delta_t|^2)\rho(\mathbf{x} + \mathbf{e}_i\delta_t)\mathbf{e}_i\delta_t = \frac{\mathbf{F}}{G^2\delta_x^2\rho}, \\ \nabla p_{\mathrm{LBE}} &= \nabla\left(p_{\mathrm{EOS}} + \frac{G^2\delta_x^2}{2}\rho^2\right) = \left(\frac{dp_{\mathrm{EOS}}}{d\rho} + G^2\delta_x^2\rho\right)\frac{\mathbf{F}}{G^2\delta_x^2\rho}.\end{aligned} \tag{61}$$

From the above discussion, it can be seen that the present collision process is actually easy to implement, and all the involved computations can be locally performed except for the simple interaction force. Therefore, compared with

the pseudopotential LB model, known as the simplest multiphase LB model, no additional computational complexity is introduced in the present LB model. Based on our numerical test, the computational cost of the present model is only 1.178 times as much as that of a pseudopotential MRT LB model by Li *et al.* [5].

## V. VALIDATIONS

In the following numerical validations, the basic simulation parameters are set as $\varpi = 1/6$, $a = 1$, $b = 4$, $R = 1$, and $\delta_x = 1$. Firstly, one-dimensional flat interface is simulated to validate the relation $W \propto K_{\text{INT}}$. Here, the interface thickness is measured from $\rho = 0.95\rho_g + 0.05\rho_l$ to $0.05\rho_g + 0.95\rho_l$, with $\rho_g$ and $\rho_l$ denoting the coexistence gas and liquid densities, respectively. Fig. 1 shows the variation of $W$ with $K_{\text{INT}}$ for the reduced temperature $T_r = T/T_c$ being 0.9, 0.8, 0.7, 0.6, and 0.5, and the scaling factor $K_{\text{EOS}}$ being 1. Good proportionality between $W$ and $K_{\text{INT}}$ can be clearly seen, and the proportionality constants (i.e., the slopes of the dashed lines) are 6.3682, 4.3490, 3.4403, 2.8787, and 2.5178 for $T_r = 0.9$, 0.8, 0.7, 0.6, and 0.5, respectively, which agree very well with the corresponding analytical results (6.3711, 4.3574, 3.4423, 2.8875, and 2.5052 for $T_r = 0.9$, 0.8, 0.7, 0.6, and 0.5, respectively) predicted by the pressure tensor. Note that when $T_r$ is relatively small (such as $T_r = 0.5$), $K_{\text{INT}}$ should be increased to widen the interfacial region for numerical stability in the simulation. Fig. 2 shows the density profile across interface for $K_{\text{EOS}} = 10^{-2}$, $10^{-1}$, $10^0$, $10^1$, and $10^2$, and $T_r = 0.8$. Here, the interface thickness is fixed at $W = 10$ (i.e., $K_{\text{INT}} = 2.2949$). It can be seen that the density profiles for different $K_{\text{EOS}}$ are identical, which indicates that both the coexistence densities and interface thickness are not affected by $K_{\text{EOS}}$. Moreover, the density profile obtained by simulation is in good agreement with the analytical one predicted by the pressure tensor, which further validates the present LB model for multiphase flows.

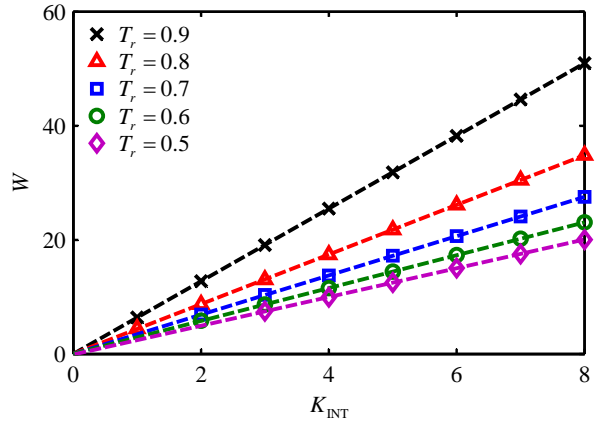

FIG. 1. Variation of the interface thickness $W$ with the scaling factor $K_{\text{INT}}$. The reduced temperature $T_r$ is chosen as 0.9, 0.8, 0.7, 0.6, and 0.5, respectively, and the scaling factor $K_{\text{EOS}}$ is fixed at 1. The dashed lines, obtained by linear fitting in a least-squares sense, are given as $6.3682 K_{\text{INT}} + 0.0156$, $4.3490 K_{\text{INT}} + 0.0392$, $3.4403 K_{\text{INT}} + 0.0218$, $2.8787 K_{\text{INT}} + 0.0527$, and $2.5178 K_{\text{INT}} - 0.0829$ for $T_r = 0.9$, 0.8, 0.7, 0.6, and 0.5, respectively.

Numerical simulation of two-dimensional circular droplet is then carried out to validate the relation $\sigma \propto K_{\text{EOS}} K_{\text{INT}}$. Here, the surface tension is computed via Laplace's law, i.e., $\sigma = (p_{\text{in}} - p_{\text{out}}) D/2$ with $p_{\text{in}}$ and $p_{\text{out}}$ denoting the pressure inside and outside the droplet, respectively, and $D$ being the droplet diameter measured at $\rho = (\rho_g + \rho_l)/2$. Fig. 3 shows the variation of $\sigma$ with $K_{\text{INT}}$ for $T_r = 0.9$, 0.8, 0.7, 0.6, and 0.5, and the scaling factor $K_{\text{EOS}} = 1$. As seen, $\sigma$ is indeed proportional to $K_{\text{INT}}$. The proportionality constants are 0.003157, 0.009115, 0.01716, 0.02725, and 0.03979 for $T_r = 0.9$, 0.8, 0.7, 0.6, and 0.5, respectively, which are in good agreement with the corresponding analytical results (0.003146, 0.009081, 0.01710, 0.02716, and 0.03954 for $T_r = 0.9$, 0.8, 0.7, 0.6, and 0.5, respectively) predicted by the pressure tensor. Compared with the simulation of flat interface, $K_{\text{INT}}$ required for numerical stability significantly increases when $T_r = 0.5$. Fig. 4 shows the variation of $\sigma$ with $K_{\text{EOS}}$ for various $T_r$. Here, the interface thickness is fixed at $W = 20$ for $T_r \geq 0.6$ and 30 for $T_r = 0.5$. The results demonstrate the perfect proportionality between $\sigma$ and $K_{\text{EOS}}$, and also suggest that $\sigma$ can be adjusted in a wide range via $K_{\text{EOS}}$.

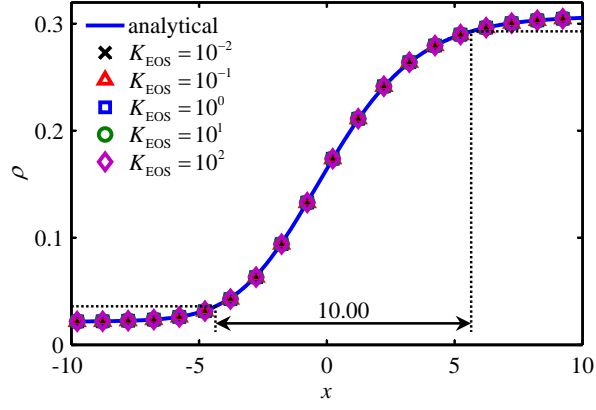

FIG. 2. Density profiles across interface obtained with various $K_{\rm EOS}$. The reduced temperature $T_r$ and interface thickness $W$ are fixed at 0.8 and 10, respectively. The solid line is the analytical profile predicted by pressure tensor. The dotted lines depict the interfacial region defined from $\rho = 0.95\rho_g + 0.05\rho_l$ to $0.05\rho_g + 0.95\rho_l$.

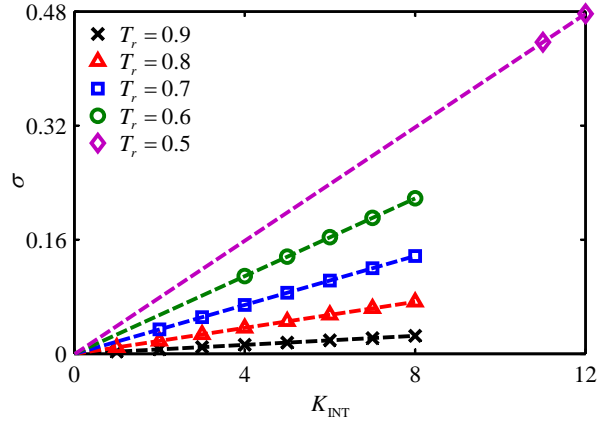

FIG. 3. Variation of the surface tension $\sigma$ with the scaling factor $K_{\rm INT}$. The reduced temperature $T_r$ is chosen as 0.9, 0.8, 0.7, 0.6, and 0.5, respectively, and the scaling factor $K_{\rm EOS}$ is fixed at 1. The dashed lines, obtained by linear fitting in a least-squares sense, are given as $0.003157 K_{\rm INT} - 3.060 \times 10^{-5}$, $0.009115 K_{\rm INT} - 8.310 \times 10^{-5}$, $0.01716 K_{\rm INT} - 8.398 \times 10^{-5}$, $0.02725 K_{\rm INT} - 2.251 \times 10^{-5}$, and $0.03979 K_{\rm INT} - 7.627 \times 10^{-4}$ for $T_r =$ 0.9, 0.8, 0.7, 0.6, and 0.5, respectively.

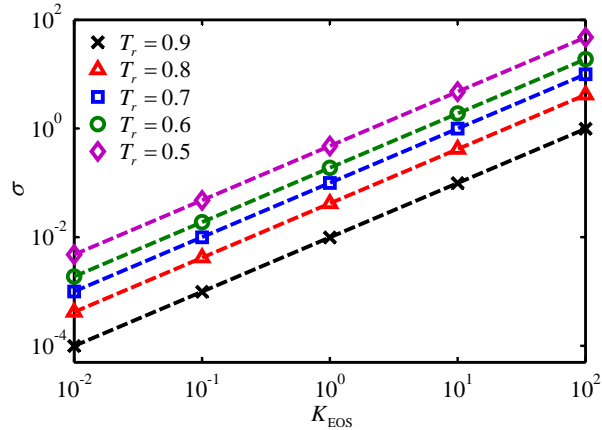

FIG. 4. Variation of the surface tension $\sigma$ with the scaling factor $K_{\rm EOS}$. The reduced temperature $T_r$ is chosen as 0.9, 0.8, 0.7, 0.6, and 0.5, respectively, and the interface thickness $W$ is fixed at 20 for $T_r \geq 0.6$ and 30 for $T_r = 0.5$, respectively. The dashed lines, obtained by linear fitting in a least-squares sense, are given as $0.009879 K_{\rm EOS} + 3.570 \times 10^{-17}$, $0.04174 K_{\rm EOS} + 1.428 \times 10^{-16}$, $0.09959 K_{\rm EOS} + 3.427 \times 10^{-16}$, $0.1887 K_{\rm EOS} + 6.283 \times 10^{-16}$, and $0.4758 K_{\rm EOS} + 1.059 \times 10^{-7}$ for $T_r =$ 0.9, 0.8, 0.7, 0.6, and 0.5, respectively.

[1] Y. H. Qian, D. d'Humières, and P. Lallemand, Europhys. Lett. **17**, 479 (1992).
[2] P. Lallemand and L.-S. Luo, Phys. Rev. E **61**, 6546 (2000).
[3] S. Chapman and T. G. Cowling, *The Mathematical Theory of Non-Uniform Gases*, 3rd ed. (Cambridge University Press, Cambridge, 1970).
[4] R. Huang and H. Wu, J. Comput. Phys. **327**, 121 (2016).
[5] Q. Li, K. H. Luo, and X. J. Li, Phys. Rev. E **87**, 053301 (2013).



\end{document}